\documentclass[12pt]{iopart}
\usepackage{graphicx}
\DeclareGraphicsExtensions{.eps}
\usepackage[english]{babel}
\usepackage{amsfonts}
\usepackage{amssymb}
\expandafter\let\csname equation*\endcsname\relax
\expandafter\let\csname endequation*\endcsname\relax
\usepackage{amsmath}
\usepackage[ansinew]{inputenc}
\usepackage{multirow}
\usepackage{mathtools}
\usepackage{wasysym}
\usepackage{color}
\usepackage{verbatim}
\usepackage{mcite}
\usepackage{array}

\begin{document}

\title{Scaling and universality in nonlinear optical quantum graphs containing star motifs}

\author{R Lytel$^1$, S Shafei$^1$, M G Kuzyk$^1$}

\address{$^1$ Department of Physics and Astronomy, Washington State University, Pullman,
Washington, US}

\begin{abstract}
Quantum graphs have recently emerged as models of nonlinear optical, quantum confined systems with exquisite topological sensitivity and the potential for predicting structures with an intrinsic, off-resonance response approaching the fundamental limit.  Loop topologies have modest responses, while bent wires have larger responses, even when the bent wire and loop geometries are identical. Topological enhancement of the nonlinear response of quantum graphs is even greater for star graphs, for which the first hyperpolarizability can exceed half the fundamental limit. In this paper, we investigate the nonlinear optical properties of quantum graphs with the star vertex topology, introduce motifs and develop new methods for computing the spectra of composite graphs.  We show that this class of graphs consistently produces intrinsic optical nonlinearities near the limits predicted by potential optimization.  All graphs of this type have universal behavior for the scaling of their spectra and transition moments as the nonlinearities approach the fundamental limit.
\end{abstract}


\submitto{\NJP}

\maketitle

\section{Introduction}

Nonlinear optical materials are quantum systems with polarizabilities that are nonlinear functions of external electromagnetic fields.  Harmonic generation \cite{maker65.02,bloem68.01,bass69.01}, electro-optics \cite{wayna00.01}, saturable absorption \cite{tutt93.01} , phase conjugation \cite{yariv77.01,yariv78.01}, four-wave mixing \cite{boyd92.01,lytel86.02}, optical bistability \cite{winfu80.01,gibbs84.01}, ultrafast optics \cite{weine11.01}, and waveguide switching \cite{lytel84.01,vanec91.01} are among the many processes in NLO materials \cite{zyss85.01,boyd09.01,kuzyk10.01} of interest in communications, instrumentation, networking, image processing, and many other fields \cite{horna92.01,kuzyk06.06}.

The polarization vector for a general system is a complex function of every allowed transition moment for the material, including electronic, vibrational, and rotational transitions, and their corresponding transition energies and damping factors.  Off-resonance, the polarization vector in most materials is dominated by weak but ultrafast electronic transitions and is a power series in the contractions of the nth order susceptibility tensor with n-1 field components, starting with the linear polarization term $\alpha_{ij}E_{j}$ and followed by the first hyperpolarizability term $\beta_{ijk}E_jE_k$, the second hyperpolarizability term $\gamma_{ijkl}E_jE_kE_l$ and so on (sum on repeated indices is implied).

For bulk materials, the polarization expansion provides a means to measure the symmetry properties of the susceptibilities and their bulk values.  On the molecular level, the expansion describes the off-resonance response of a single molecule to external optical fields.  The hyperpolarizability tensors become fully symmetric and their global properties are set by the characteristics of the states and their spectra.  These are manifestations of the topology of the system, and their contributions to the susceptibilities vary over specific ranges with values that depend on the geometry of the system.

Scale-free, \emph{intrinsic} hyperpolarizability tensors in the off-resonance regime may be created by normalizing them to a maximum value determined by fundamental limits set by the generalized Thomas-Reiche-Kuhn (TRK) sum rules,\cite{kuzyk00.01} which constrain the sums of oscillator strengths responsible for the nonlinear response \cite{kuzyk00.01,kuzyk09.01,kuzyk06.03,kuzyk03.01}.  Hyperpolarizabilities normalized this way enable direct comparisons of the intrinsic response without regard to size.

Quantum systems have yet to be found that achieve the maximum allowed values of the response, though searches for optimized potentials have led to molecules with record hyperpolarizabilities \cite{zhou06.01,zhou07.02,perez07.01,perez09.01,ather12.01,burke12.01}, still well short of the fundamental limits. Monte Carlo simulations of the entire space of allowed states and spectra consistent with the TRK sum rules \cite{bello08.01,wang99.01,kuzyk06.01} have proved the existence of solutions that approach the maximum values \cite{kuzyk08.01,shafe10.01}, though the nature of the Hamiltonian that can produce these is still unknown. Detailed analysis of the optimized systems reveals that they share certain universal properties at their global maxima \cite{watkins09.01}.

A quantitative rule of thumb, the so-called Three-Level Ansatz, emerged from the theory of fundamental limits (TFL) and states that only three states contribute for a system with a nonlinearity close to the limit, consistent with all observations and analysis to date.  This strongly suggests that systems in which many states contribute yield low hyperpolarizabilities \cite{kuzyk00.01}, also consistent with all known data.  Conversely, models with states and spectra mimicking three-level systems, with large energy gaps and small transition moments of the higher-lying excited states \cite{shafe10.01}, such as a one-dimensional, particle in a box-like spectrum, where $E_n \propto n^2$, can potentially lead to large hyperpolarizabilities.  Quantum nanowires and molecular systems with tight confinement possess states and spectra with such properties and have motivated the present work.

A quantum graph (QG) is a general confinement model for quasi-one dimensional electron dynamics.  The generalized QG model of an $N$ electron structure constrains dynamics to the edges of a metric graph.  Dynamics are governed by a self-adjoint Hamiltonian with a complete set of eigenstates and eigenvalues.  The general Hamiltonian contains momentum, position, and spin operators, as well as functions of each describing particle-particle interactions, band spectra, coupling to external fields, and other interactions.  Transitions between states, described by a set of transition moments, determine the response of the graph to an external optical field.  The canonical commutation relations guarantee that the TRK sum rules hold for the transition moments, providing constraints and relations among the various allowed transitions in the system \cite{kuzyk00.01,kuzyk06.03}.

The one-electron version of the generalized quantum graph model (hereafter referred to as the \emph{elementary QG}) is an exactly solvable model of quantum chaos and related phenomena.  Quantum graphs with zero potential energy (bare edges) and nonzero potentials (dressed edges) have been solved using periodic orbit theory and extensively studied for their statistical properties and energy spectra \cite{kotto97.01,kotto99.02,blume02.01,blume02.02,dabag04.01,dabag02.01,dabag03.01,dabag07.01,gnutz10.01}.  This model has recently been applied to calculate the off-resonance first ($\beta_{ijk}$) and second ($\gamma_{ijkl}$) hyperpolarizability tensors (normalized to their maximum values) of elementary graphical structures, such as wires, closed loops, and star vertices \cite{lytel12.01,shafe12.01} and to investigate the relationship between the topology and geometry of a graph and its nonlinear optical response \cite{lytel13.01} through its hyperpolarizability tensors.  The results showed that the elementary QG model of a 3-edge star graph generated a first hyperpolarizability over half the fundamental limit and a second hyperpolarizability whose range was between 20-40 percent of the fundamental limit.  The elementary QG model also reproduced universal scaling results predicted by the TFL for a general quantum system, suggesting it is a sound model for exploring tightly confined nonlinear optical systems with electron dynamics on the edges of a quantum graph.

The present work significantly expands the analysis of the elementary QG in nonlinear optics to graphs containing many star vertices in order to identify graphs with large intrinsic nonlinearities, explore methods for solving graphs containing arbitrary numbers of stars, and determine their universal properties as they approach their optimum values.  Despite its simplicity, the elementary QG model reflects universal scaling behavior of quantum dynamical systems when the nonlinearities approach their maximum intrinsic values.  As such, explorations of the limits obtainable from the model may prove useful to designers of nanowires and networks and form the foundation of a comprehensive analysis of the nonlinear optical response of quantum confined systems whereby the edge potentials (dressing), boundary conditions (topology), and multi-dimensionality (including fractals) represent the molecular physics of a quantum wire system.

Section \ref{sec:QGreview} reviews the calculation of the first and second hyperpolarizability tensors for all of the graphs in table {\ref{tab:resultsTable}, once the graph has been solved for its eigenstates and energy spectrum.  Section \ref{sec:motif} displays the states and spectra for star graphs and introduces the analytical star \emph{motif} for evaluating quantum graphs comprised of arbitrary numbers of wires, loops, and stars.  Section \ref{sec:motif} shows how to use the motifs to derive the states and spectra for graphs in the table, as well as for more complex graphs.  Section \ref{sec:results} discusses the intrinsic limits of the classes of graphs exhibited in table \ref{tab:resultsTable} and presents their scaling properties as the fundamental limits of each graph are approached.  It is also shown how complex quantum graphs might be able to achieve record intrinsic nonlinearities due to the tunability of the level spacings made possible by multiple sets of secular equations for motifs comprising a graph.  Section \ref{sec:outlook} summarizes the application of the elementary QG model to elementary and composite graphs and points to the next direction, dressed quantum graphs with multiple electron dynamics.  Several appendices provide details of the computations for the graphs and explain the handling of degeneracies and other details.

Two new and fundamental results emerge from this work that can aide the molecular designer of nanowire and quantum-confined systems for nonlinear optics.  The first is that the global properties of structures comprised of star motifs are nearly identical as the geometry of the structure is tuned toward maximum values, hinting at the same universal properties observed in other studies. The second is that these topologies generally have the largest intrinsic responses achievable to date, and might be realized in quasi-one dimensional nanostructures.

\begin{table}\centering\scriptsize
\caption{Intrinsic nonlinearities of topological classes of quantum graphs. The first ($\beta_{xxx}$) and second ($\gamma_{xxxx}$) hyperpolarizabilities shown are the largest values for the geometries within the specific topological class.  The first hyperpolarizability tensor norm $\beta_{norm}$ is defined and calculated in the text, and is an invariant for the topological class. In all cases except closed loops, the maximum value of $\beta_{xxx}$ is equal to $\beta_{norm}$, indicating that the topology allows the graph to assume its best configuration for the xxx component, which usually means that the yyy component vanishes.  Loops are so tightly constrained that it is impossible for a loop to have one of its diagonal components at zero when the other is nonzero.\label{tab:resultsTable}}

\newcolumntype{S}{>{\centering\arraybackslash} m{2cm} } 
\begin{tabular}{S S S S S S }
  \hline\hline
Graph  & Geometry & Topology & $\beta_{norm}$ & $\left| \beta_{xxx} \right|$ & $\gamma_{xxxx}$ \\
  \hline\hline
\\
\includegraphics{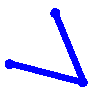} & bent wire & line & 0.172 & 0.172 & -0.126 to 0.007 \\
\hline
\\
\includegraphics{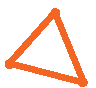} & triangle & loop & 0.086 & 0.049 & -0.138 to 0 \\
\hline
\\
\includegraphics{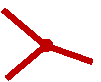} & 3-star & 3-fork & 0.58 & 0.58 & -0.138 to 0.3 \\
\includegraphics{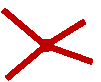} & 4-star & 4-fork & 0.53 & 0.53 & -0.125 to 0.27 \\
\includegraphics{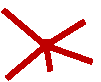} & 5-star & 5-fork & 0.51 & 0.51 & -0.11 to 0.26 \\
\includegraphics{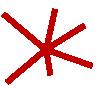} & 6-star & 6-fork & 0.51 & 0.51 & -0.11 to 0.26 \\
\includegraphics{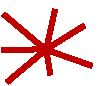} & 7-star & 7-fork & 0.51 & 0.51 & -0.11 to 0.26 \\
\hline
\\
\includegraphics{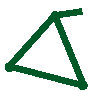} & lollipop & star-loop & 0.62 & 0.62 & -0.12 to 0.20 \\
\includegraphics{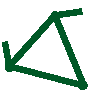} & bull & star-loop & 0.53 & 0.53 & -0.09 to 0.20 \\
\includegraphics{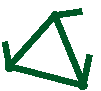} & lollipop bull & star-loop & 0.51 & 0.51 & -0.09 to 0.19 \\
\includegraphics{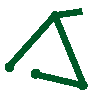} & lollipop & 3-fork & 0.33 & 0.33 & -0.11 to 0.13 \\
\includegraphics{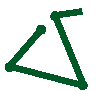} & lollipop & line & 0.17 & 0.17 & -0.09 to 0.006 \\
\hline
\\
\includegraphics{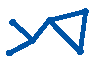} & barbell & 2-fork lollipop & 0.54 & 0.54 & -0.104 to 0.214 \\
\includegraphics{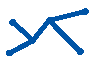} & barbell & dual 2-fork & 0.43 & 0.43 & -0.13 to 0.22 \\
\includegraphics{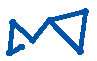} & barbell & star-loop & 0.41 & 0.41 & -0.07 to 0.11 \\
\includegraphics{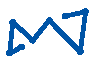} & barbell & line & 0.14 & 0.14 & -0.085 to 0.006 \\
\includegraphics{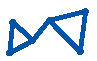} & barbell & loop & 0.11 & 0.11 & -0.1 to 0.002 \\
\hline\hline
\end{tabular}
\end{table}

\section{Nonlinear optics in the elementary QG model}\label{sec:QGreview}

The dynamics of an electron on a quantum graph are described by a self-adjoint Hamiltonian operating on the edges of the graph, with complex amplitude and probability conservation (hereafter referred to as flux conservation throughout the paper) at all internal vertices and fixed, infinite potentials at the termination vertices (where the amplitude vanishes).  The physics of the eigenstates and their spectra have been previously described, along with a suitable lexicography for describing the \emph{union} operation for creating eigenstates that are complete and orthonormal from the edge functions that solve the equations of motion for the same Hamiltonian, with the same eigenvalues, on each edge \cite{shafe12.01}.

The graph is specified by the location of its vertices and the edges connecting the vertices.  A set of vertices with arbitrary locations in the 2D plane but fixed connections specifies a topological class of graphs.  For a fixed topology, the variation of vertex locations specifies various geometries for the graph.  Since motion is confined to the graph edges and is continuous at each vertex, the energy spectrum depends only on the edge lengths and the boundary conditions, ie, the topology.  The lengths and angular positions of the edges determine the projections of electron motion onto a fixed, external reference axis.  The projections summed over all edges yield the transition moments required to compute the tensor elements of the hyperpolarizabilities.  Regardless of how the axes used to define the vertices are chosen, the various tensor components may be used to assemble any component in a different frame through the rotation group.

The study of the nonlinear optical properties of a specific graph topology requires solving the graph for its eigenstates and spectra as functions of its edge lengths and using them to compute a set of transition moments for the graph from which the hyperpolarizability tensors may be computed. We first review how the tensors are calculated, as the method is universal for any graph.  Then we show how to calculate the states and spectra for the graphs.

\subsection{Hyperpolarizability tensors}\label{tensorCalcs}

In this paper, we focus our analysis on the off-resonance, nonlinear first ($\beta_{ijk}$) and second ($\gamma_{ijkl}$) hyperpolarizabilities.  In this regime, both tensors are fully symmetric, with four nonzero components for $\beta_{ijk}$ and five nonzero components for $\gamma_{ijkl}$.  Scale-independent (intrinsic) tensors are created by normalizing each tensor to the fundamental limit.  The fundamental limits are the highest attainable first and second hyperpolarizabilities and depend on the number of electrons, $N$, and the energy gap between the ground and the first excited state, $E_{10}$. They are given by \cite{kuzyk00.01, kuzyk00.02}
\begin{equation}\label{sh-betaMax}
\beta_{max} = 3^{1/4} \left(\frac{e\hbar}{m^{1/2}}\right)^3 \frac{N^{3/2}}{E_{10}^{7/2}}
\end{equation}
and
\begin{equation}\label{sh-gammaMax}
\gamma_{max} = 4 \left(\frac{e^4\hbar^4}{m^2}\right) \frac{N^{2}}{E_{10}^{5}} .
\end{equation}
Throughout this paper, all tensor components of the hyperpolarizabilities are normalized by these maxima, ie,
\begin{equation}\label{IntrinsicBetaGamma}
\gamma_{ijkl} \rightarrow \frac {\gamma_{ijkl}} {\gamma_{max}} \hspace{2em} \beta_{ijkl} \rightarrow \frac {\beta_{ijkl}} {\beta_{max}} .
\end{equation}
The second hyperpolarizability normalized this way has a largest negative value equal to $-(1/4)$ of the maximum value.  The first intrinsic hyperpolarizability tensor for 2D graphs may then be written as
\begin{eqnarray}\label{betaInt}
\beta_{ijk} &\equiv& \frac{\beta}{\beta_{max}} = \left(\frac{3}{4}\right)^{3/4} {\sum_{n,m}}' \frac{\xi_{0n}^{i}\bar{\xi}_{nm}^{j}\xi_{m0}^{k}}{e_n e_m} ,
\end{eqnarray}
where $\xi_{nm}^{i}$ and $e_n$ are normalized transition moments and energies, defined by
\begin{equation}\label{xNMnorm}
\xi_{nm}^{i} = \frac{r_{nm}^{i}}{r_{01}^{max}}, \qquad e_{n} = \frac{E_{n0}}{E_{10}},
\end{equation}
with $r^{i=1}=x$ and $r^{i=2}=y$, and where
\begin{equation}\label{Xmax}
r_{01}^{max} = \left(\frac{\hbar^2}{2 m E_{10}}\right)^{1/2}.
\end{equation}
$r_{01}^{max}$ represents the largest possible transition moment value of $r_{01}$.  According to Eq. (\ref{xNMnorm}), $e_0 = 0$ and $e_1 = 1$. $\beta_{ijk}$ is scale-invariant and can be used to compare molecules of different shapes and sizes.
Similarly, the second intrinsic hyperpolarizability is given by
\begin{eqnarray}\label{gammaInt}
\gamma_{ijkl} &=& \frac{1}{4} \left({\sum_{n,m,l}}' \frac{\xi_{0n}^{i}\bar{\xi}_{nm}^{j}\bar{\xi}_{ml}^{k}\xi_{l0}^{l}}{e_n e_m e_l} - {\sum_{n,m}}' \frac{\xi_{0n}^{i}\xi_{n0}^{j}\xi_{0m}^{j}\xi_{m0}^{k}}{e_n^2 e_m}\right) . \nonumber \\
\end{eqnarray}

A molecular designer is most interested in knowing which geometries of a given topology yield a larger response, and which topologies show the most promise for enabling a specific geometry to have one of the larger possible responses.  This knowledge is obtained by specifying a fixed topology, such as any of those in table \ref{tab:resultsTable}, and calculating the response for a large number of possible geometries in order to discover the \emph{best} shape.  By best, the experimentalist usually means the one with the largest value of the hyperpolarizability in a lab frame whose $x$-axis is known and usually used to reference the optical field polarizations interacting with the material.

The specification of a graph through its vertices, the calculation of its states and spectra, and the sampling of large numbers of geometries to create ensembles of transition moments, energies, and hyperpolarizabilities, is a Monte Carlo computation.  The results of such a calculation are a set of tensors for a topological class of graphs whose variability is solely determined by the geometrical properties of the graphs.

Using the rotation properties of the tensors, it is straightforward to identify the preferred diagonal orientation for any specified graph \cite{lytel13.01}, the one for which the hyperpolarizability along a specific axis is maximum.  It is typical for $\beta_{xxx}$ and $\gamma_{xxxx}$ to be largest along different axes, so the preferred diagonal orientation of each may be and usually is different.

From here on, we denote $\beta_{xxx}$ ($\gamma_{xxxx}$) as the largest diagonal tensor component of the intrinsic first (second) hyperpolarizability when the graph is in its preferred diagonal frame.  The Monte Carlo method, applied to each graph in an ensemble calculation, uses the tensor components it computes to find the angle through which the graph should be rotated to maximize its laboratory component.  In this way, scale-invariant, orientation-independent intrinsic hyperpolarizabilities for topological classes of graphs may be studied as functions of graph geometry.

For a reference frame that is rotated $\phi$ degrees with respect to the initial reference frame, the diagonal components, $\beta_{xxx}(\phi)$ and $\gamma_{xxxx}(\phi)$, can be determined using
\begin{eqnarray}\label{BetaCartesian}
\beta_{xxx}(\phi) &=& \beta_{xxx}\cos^3\phi \nonumber + 3\beta_{xxy}\cos^2\phi \sin\phi \nonumber \\ &+&  3\beta_{xyy}\cos\phi \sin^2\phi + \beta_{yyy}\sin^3\phi,
\end{eqnarray}
and
\begin{eqnarray}\label{GammaCartesian}
\gamma_{xxxx}(\phi) &=& \gamma_{xxxx}\cos^4\phi+4\gamma_{xxxy}\cos^3\phi\sin\phi + 6\gamma_{xxyy}\cos^2\phi \sin^2\phi\nonumber \\
&+& 4\gamma_{xyyy}\cos\phi \sin^3\phi + \gamma_{yyyy}\sin^4\phi
\end{eqnarray}
where the value of $\phi$ that maximizes the left-hand side of either equation is usually different for each of $\beta_{xxx}$ and $\gamma_{xxxx}$, and the tensor components on the right-hand side of either equation are referenced to the graph's intrinsic frame where the hyperpolarizabilities are calculated.  By definition, $\beta_{xxx}$ ($\gamma_{xxxx}$) is at an extreme value when the graph is rotated through $\phi$.  Once the graph is solved and the tensor components are known in its frame, $\phi$ is easily found by maximizing (\ref{BetaCartesian}) for $\beta_{xxx}$ and (\ref{GammaCartesian}) for $\gamma_{xxxx}$.  The tensor norms are invariant under any transformation and provide immediate insight into the limiting responses of the graphs.  They are given by
\begin{equation}\label{BetaNorm}
|\beta| = \left( \beta_{xxx}^2+3\beta_{xxy}^2+ 3\beta_{xyy}^2+\beta_{yyy}^2\right)^{1/2}
\end{equation}
and
\begin{equation}\label{GammaNorm}
|\gamma| = \left( \gamma_{xxxx}^2+4\gamma_{xxxy}^2+ 6\gamma_{xxyy}^2+4\gamma_{xyyy}+\gamma_{yyyy}^2\right)^{1/2}
\end{equation}
These are the magnitudes of the graph's hyperpolarizabilities and are both scale and orientation-independent. The use of tensors to extract the nonlinear optical response as a function of geometry and topology is most easily achieved by transforming the Cartesian tensors to spherical tensors. The transformation from a Cartesian to a spherical tensor representation is achieved using Clebsch-Gordon coefficients and has been extensively discussed in the literature \cite{jerph78.01,bance10.01}, as has their application to aromatic systems \cite{joffr92.01} and quantum graphs \cite{lytel13.01}.

\subsection{States and spectra}\label{statesAndSpectra}

To calculate the hyperpolarizabilities of a graph, such as that depicted in figure \ref{fig:graphNEW}, the graph must be solved for its eigenstates and energy spectrum.  The eigenstates are unions of the edge functions over the graph \cite{shafe12.01}:
\begin{equation}\label{eigenFunction}
\psi_n(s)= \cup_{i=1}^{E} \phi_n^{i}(s_{i})
\end{equation}
and the edge functions for an edge connecting a vertex with amplitude $A_{n}$ to vertex with amplitude $B_{n}$ may be written in a canonical form that automatically matches (nonzero) amplitudes at each internal vertex:
\begin{equation}\label{edgeFunctionAB}
\phi_n^{i}(s_i)= \frac {A_{n}^{(i)} \sin k_{n}(a_i-s_i)+B_{n}^{(i)}\sin k_{n}s_i}{\sin k_{n}a_i}
\end{equation}
Vertices with zero amplitude occur when the edges in the graph are rationally-related and are discussed in the Appendix.  For the rest of this paper, we assume the edges are irrationally-related, so that the denominator in (\ref{edgeFunctionAB}) never vanishes.
\begin{figure}\centering
\includegraphics[width=90mm]{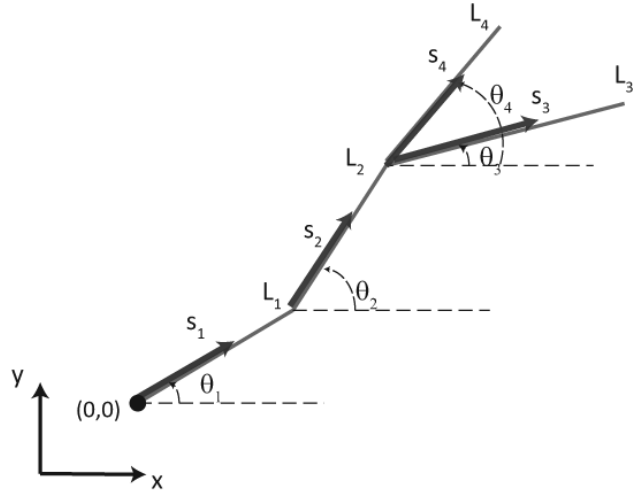}
\caption{A four-edge quantum graph.  Each edge has its own longitudinal coordinate $s_{i}$ ranging from zero to $L_{i}$.  The projection $x(s_{i})$ of an edge onto the $x-axis$ is measured from the origin of the coordinate system attached to the graph (and not to the beginning of the edge itself).  For example, $x(s_1)=s_1\cos\theta_1$ while $x(s_2)=L_1\cos\theta_1 + s_2\cos\theta_2$ and so on.}\label{fig:graphNEW}
\end{figure}

Terminal vertices with zero amplitude will take the form of \ref{edgeFunctionAB} with one of $A_{n}$ or $B_{n}$ equal to zero.  A graph with $N_{V}$ internal vertices and $N_{v}$ external vertices generates exactly $N_{V}$ flux-conserving equations for the set of $N_{V}$ internal amplitudes.  Solutions exist only if the determinant of the matrix of coefficients of the $N_{V}$ coupled amplitude equations vanishes.  This condition produces the secular or characteristic equation for the graph and determines the exact energy spectrum.  Since the boundary conditions in the elementary QG model are independent of the angles the edges make with respect to one another, the secular equation is independent of angles and depends only on dimensionless parameters $k_{n}a_{i}$.  For a given configuration of vertices, the distance between them and the rules by which they are connected, i.e., the topology of the graph, determines the energy spectrum.  Graphs with identical geometries, such as those shown in the sets in Table \ref{tab:resultsTable} have energy spectra and hence eigenstates that differ due to their topological differences.

Except for bent wires and closed loops, the secular equation of a graph is generally a transcendental equation.  Exact solutions of the form $k_{n}=f(n)$ have been produced using periodic orbit theory \cite{blume02.01,blume02.02,dabag04.01,dabag03.01}.  Accurate solutions are easily found numerically without resorting to periodic orbits, however.  From these, the internal amplitudes $A_{n}^{(i)}$ and $B_{n}^{(i)}$ may be calculated relative to the same normalization constant.  Normalizing the eigenfunction produces the states required to compute the transition moments, and these and the energies may be used in (\ref{betaInt}) and (\ref{gammaInt}) to compute the intrinsic hyperpolarizability tensors.

It should be noted that the transition moments are sums (not unions) over edges of the following form:
\begin{equation}\label{xNM}
x_{nm}=\sum_{i=1}^{E}\int_{0}^{a_{i}}\phi_{n}^{*i}(s_{i})\phi_{m}^{i}(s_{i})\ x(s_{i})ds_{i}
\end{equation}
where $\phi_{m}^{i}(s_{i})$ are the normalized edge wave functions whose coefficients and normalization were just calculated for the graph and $x(s_{i})$ is the $x-component$ of $s_{i}$, measured from the origin of the graph (and not of the edge), and is a function of the prior edge lengths and angles.  With edge wave functions of the form of (\ref{edgeFunctionAB}), the computation of the transition moments requires integrals of products of sines and cosines with either $s$ or $1$, all of which are calculable in closed form.  Detailed examples for loops, wires, and stars are available in the literature \cite{lytel12.01,shafe12.01,lytel13.01} from our prior work.

The collective tensor properties contain all physical information about the response of every geometric configuration of a specific graph topology.  A large sampling of the space of possible geometries may be explored using a Monte Carlo calculation. The process for calculating the hyperpolarizability tensors of elementary QG's can be summarized as follows: (1)  select a particular graph topology, specifying the number of vertices and the connecting edges, (2)  generate a random set of vertices, and calculate the lengths of the edges and the angles each makes with the $x$-axis of the graph's coordinate system, (3) solve the Schr\"{o}dinger Equation on each edge of the graph, and (4) match boundary conditions at the vertices and terminal points.  This results in a set of equations for the amplitudes of the wavefunctions on each edge.  The solvability of this set requires that the determinant of the amplitude coefficients vanishes, leading to a secular equation for the eigenvalues.  The transition moments $x_{nm}$ and energies $E_{n}=\hbar^{2}k_{n}^{2}/2m$ may be used to compute the first and second hyperpolarizabilities of any graph specified by a set of vertices.

The hyperpolarizability tensors were calculated for each graph in table \ref{tab:resultsTable} once the graph was solved for its edge wavefunction amplitudes and energies.  Bent wires, loops, and stars have been solved and the hyperpolarizabilities calculated by matching boundary conditions at each vertex.  All of the graphs in the table may be solved this way, but an efficient way to solve any graph is to use motifs, as discussed next.

\section{Solving graphs using motifs}\label{sec:motif}

Connected composite graphs may be constructed from the elemental graphs, or \emph{motifs}.  The spectra of connected graphs are the solutions to their secular equations, which always take the form of combinations of the secular functions of simpler graphs.  The motifs in figure \ref{fig:motifGraphs} are sufficient to compute the states and spectra for all graphs in table \ref{tab:resultsTable}.  This section presents the motifs of the star and lollipop graphs and shows how to use them to compute a more complex graph, such as the bullgraph in the table. The remaining calculations for all of the graphs are presented in the Appendices.

The use of motifs for solving quantum graphs is a powerful tool for writing down the spectral equations almost by inspection.  We focus on graphs whose edges are irrationally-related, as this removes the possibility of degeneracies.  Degeneracies offer no difficulty \cite{pasto09.01} but also no new physics.  For completeness, the degenerate $N=3$ star graph is solved in the Appendix.  We focus here on internal vertices with degree $\leq 3$.  Scaling to stars with greater $N$ is straightforward and discussed in the Appendix, too.  Loop graphs, such as the lollipop or barbell, can have multiple sets of states, one set describing the dynamics on all edges (via the secular equation), and other, simpler sets describing dynamics on only part of the graph.  We discuss how the lollipop motif has two such sets and show how to work with this situation if it arises.  The barbell graph discussed in the Appendix has a zero energy ground state, as does any closed loop system, because there is no prong terminated at infinite potential.  As it is simple to add a state to a spectrum, this situation is also relegated to the Appendices.

\subsection{Star and lollipop graph motifs}

The nonlinearities of both the 3-star graph with edges terminated at infinite potential ($A=B=C=0$) and the lollipop with its stick terminated at infinite potential ($A=0$) have been calculated in the elementary QG model \cite{lytel13.01}. As isolated models of nonlinear, quantum confined systems, these are interesting structures because both topologies have intrinsic nonlinearities over half the fundamental limits.  These are shown in table \ref{tab:resultsTable}.  As noted above, we assume $N=3$ unless otherwise stated; the results easily generalize to $N\geq 4$.  To understand and use motifs, we need to understand the flow of flux in the star and lollipop graphs when the ends are not terminated at infinite potential.

The conservation of flux in a star graph leads to the reduced secular function $f_{star}$, where
\begin{equation}\label{reducedSecStar}
f_{star}(a_{i})=\sum_{i=1}^{E}\cot{k_{n}a_{i}}
\end{equation}
for an E-pronged star with edges $a_{i}$.

For the 3-star with edges $a,b,c$, multiply the reduced secular function by $\sin{k_{n}a} \, \sin{k_{n}b} \, \sin{k_{n}c}$, a factor that is nonzero for irrationally-related edges, and we get the secular function $F_{star}(a,b,c)$ \cite{pasto09.01,lytel13.01}
\begin{equation}\label{3starSecularF}
F_{star}(a,b,c) = \frac{1}{4}\left[\cos{k_nL_1} + \cos{k_nL_2} + \cos{k_nL_3} - 3\cos{k_nL}\right],
\end{equation}
where $L=a+b+c$, $L_1=|a+b-c|$, $L_2=|a-b+c|$, and $L_3=|a-b-c|$.
We call (\ref{3starSecularF}) the canonical form of the 3-star secular function and will use it extensively in what follows.  The combination lengths are equivalent to the edge lengths, and we freely move back and forth between them.  For example, a star graph with edges $d,e,f$ will have a secular function $F_{star}(d,e,f)$ which may be written in the form of the right hand side of Eqn (\ref{3starSecularF}) with the set $(a,b,c)$ replaced by $(d,e,f)$ in the definition of the combination lengths.

The solutions to the secular equation $F_{star}=0$ for irrational lengths have been discussed at length in Ref. \cite{pasto09.01}, where a periodic orbit expansion was derived for the eigenvalues.  They are nondegenerate and lie one to a cell between root boundaries at multiples of $\pi/L$ \cite{pasto09.01,dabag07.01}.  For our purposes, a set of solutions for any finite number of wave functions is easily found by numerically intersecting the two parts of the secular equation.  In this way, a set of nondegenerate eigenvalues may be obtained for arbitrary (but irrational) prong lengths.  Solutions may be found in Ref. \cite{lytel13.01}. The energy eigenvalues are located in cells between root boundaries \cite{pasto09.01,lytel13.01}.  Their values move around within the root boundaries but the root boundaries are fixed and scale linearly with state number.

For the lollipop graph, the secular function $F_{pop}(a,L_{tot})$ is \cite{lytel13.01}
\begin{equation}\label{lollipopSecular}
F_{pop}(a,L_{tot}) = \frac{1}{2}\left[3\cos k_n \left(a+\frac{L_{tot}}{2}\right) - \cos k_n \left(a-\frac{L_{tot}}{2}\right)\right].
\end{equation}
where $L_{tot}=b+c+d$ is the length of the loop and $a$ is the prong length.

The wavefunctions of the lollipop graph are a composite of two sets of wavefunctions, one set that is nonzero at the central vertex and on all edges, and one for wavefunctions that vanish at the origin and are exactly zero on the prong edge.  The first set correspond to the symmetric wavefunctions of a 3-sided bent wire (open at the central vertex) coupled to a nonzero prong wavefunction, while the second set correspond to the asymmetric wavefunctions of a 3-sided bent wire (open at the central vertex) with a zero prong wavefunction.  When another graph is attached to the prong, the loop-only wave functions go away and we're left with the wave functions satisfying the secular equation above.

Lollipops and 3-star graphs are in and of themselves interesting models of nonlinear optical molecular systems due to their large spatial degrees of freedom and their large intrinsic hyperpolarizabilities.  Physical systems exhibiting charge transfer paths similar to those of these graphs should also exhibit large intrinsic hyperpolarizabilities.

Consider now how to use the motifs to construct the secular functions for composite graphs, such as those in table \ref{tab:resultsTable}.  For the star graph with three terminated ends, the secular function $F_{star}$ is exactly zero.  When the ends in the motifs are unterminated, the amplitudes at the ends are nonzero and there must be flux moving in or out of these ends, since flux is conserved in the graph.  This means that the secular function is no longer zero but is related to the flux entering or leaving the unterminated vertices.  For the 3-star motif in figure \ref{fig:motifGraphs}, the canonical form of the edge functions is
\begin{eqnarray}\label{3starEdges}
\phi_n(s_{a}) &=& \frac{Z_n \sin k_n\left(a-s_{a}\right)+ A_{n}\sin k_{n}s_{a}}{\sin k_{n}a} \nonumber \\
\phi_n(s_{b}) &=& \frac{Z_n \sin k_n\left(b-s_{b}\right)+ B_{n}\sin k_{n}s_{b}}{\sin k_{n}b} \nonumber \\
\phi_n(s_{c}) &=& \frac{Z_n \sin k_n\left(c-s_{c}\right)+ C_{n}\sin k_{n}s_{c}}{\sin k_{n}c} \nonumber \\
&&
\end{eqnarray}
For unterminated ends, conservation of flux at the central vertex $Z$ produces the following secular equation relating the amplitudes at the ends and the central amplitude:
\begin{eqnarray}\label{3starSecUnterm}
Z_{n}F_{star}(a,b,c) &=& A_n\sin k_{n}b\sin k_{n}c \nonumber \\
&+& B_n\sin k_{n}a\sin k_{n}c \\
&+& C_n\sin k_{n}a\sin k_{n}b \nonumber
\end{eqnarray}
The left-hand side is the net flux through its unterminated vertices required to conserve flux at the central vertex.  If the ends are terminated, the left-hand side vanishes, reproducing the secular equation for a terminated star graph, $F_{star}(a,b,c)=0$.  For unterminated ends, (\ref{3starSecUnterm}) relates the amplitudes at the ends and at the central vertex through a single equation.

\begin{figure}\centering
\includegraphics[width=1.7in]{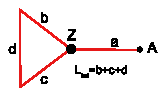}\includegraphics[width=1.7in]{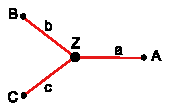}\\
\includegraphics[width=1.7in]{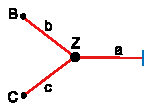}\includegraphics[width=1.7in]{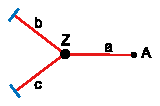}\\
\caption{The four primary motifs for constructing any graph.}
  \label{fig:motifGraphs}
\end{figure}

Similar remarks hold for the unterminated lollipop in figure \ref{fig:motifGraphs}.  The exact expression for the flux in/out of the lollipop motif is
\begin{equation}\label{lollipopSecUnterm}
Z_{n}F_{pop}(a,L_{tot}) = A_{n}\cos k_{n}L_{tot}/2.
\end{equation}
The left-hand side is the total flux exiting the central vertex Z and entering the vertex A.  When $A=0$, the flux conservation equation becomes $F_{pop}(a,L_{tot})=0$.  This determines the eigenvalues of the terminated lollipop graph where there is flux moving on all of its edges but never exiting at vertex A.  As noted above, the terminated lollipop has an additional spectrum comprised of wave functions where there is exactly zero flux on edge a at all times, i.e., flux just circulates around the loop.  This set must be included in the total spectrum of the lollipop.  We next show how to use the motifs to calculate the secular equation for a graph consisting of combined motifs.

\subsection{Bull graph}

Consider first the graph in figure \ref{fig:bullGraph}, known as a bullgraph.  Without motifs, we solve for the secular equation by writing down the edge functions for each of the four edges in the graph (treating $L_{1}+L_{2}\equiv d$ as a single edge (for purposes of computing the eigenvalues, not the transition moments), as described in \cite{lytel12.01}) that meet continuity boundary conditions at the vertices:
\begin{eqnarray}\label{edgeStates2PL}
\phi_{n}^{a}(s_{a}) &=& A_{n}\frac{\sin k_{n}\left(a-s_{a}\right)}{\sin k_{n}a}\nonumber \\
\phi_{n}^{b}(s_{b}) &=& B_{n}\frac{\sin k_{n}\left(b-s_{b}\right)}{\sin k_{n}b} \\
\phi_{n}^{c}(s_{c}) &=& B_{n}\frac{\sin k_{n}s_{c}}{\sin k_{n}c} + A_{n}\frac{\sin k_{n}(c-s_{c})}{\sin k_{n}c}\nonumber \\
\phi_{n}^{d}(s_{d}) &=& B_{n}\frac{\sin k_{n}s_{d}}{\sin k_{n}d} + A_{n}\frac{\sin k_{n}(d-s_{d})}{\sin k_{n}d}\nonumber
\end{eqnarray}

\begin{figure}\centering
\includegraphics[width=2in]{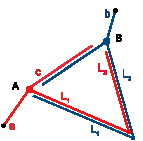}\includegraphics[width=2in]{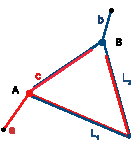}\\
\caption{The lollipop with an extra prong is a bullgraph that is topologically comprised of two 3-star motifs, each with one bent prong.  The bent prong behaves as a single wire with length equal to the sum of its two straight edges, as far as the energy spectrum is concerned.  The bend affects only the transition moments.}
\label{fig:bullGraph}
\end{figure}

Flux is conserved at both star vertices.  This yields the two equations
\begin{eqnarray}\label{fluxCons2PL}
&& A_{n}\left(\cot{k_{n}a}+\cot{k_{n}c}+\cot{k_{n}d}\right)\nonumber \\
&=& B_{n}\left(\csc{k_{n}c}+\csc{k_{n}d}\right), \\
&& B_{n}\left(\cot{k_{n}b}+\cot{k_{n}c}+\cot{k_{n}d}\right)\nonumber \\
&=& A_{n}\left(\csc{k_{n}c}+\csc{k_{n}d}\right)
\end{eqnarray}
We note immediately that the factor multiplying $A_{n}$ ($B_{n}$) in the first (third) line is the reduced secular function for a single star.  This means that the two stars will share flux via the connections, and that the flux moving across the graph is given by the second or fourth lines.  Cross-multiplying yields an equation for the wavenumbers of the graph.  This illustrates exactly how two star motifs combined to create a bullgraph. In fact, multiply the numerators of the above equations by the nonzero products $\sin{k_{n}a}\sin{k_{n}c}\sin{k_{n}d}$ for the first two and $\sin{k_{n}b}\sin{k_{n}c}\sin{k_{n}d}$ for the second two.  We then arrive at the following relationships among the two star motifs:
\begin{eqnarray}\label{fluxBull2stars}
A_{n}F_{star}(a,c,d) &=& B_{n}\sin{k_{n}a}(\sin{k_{n}c}+\sin{k_{n}d}) \\
B_{n}F_{star}(b,c,d) &=& A_{n}\sin{k_{n}b}(\sin{k_{n}c}+\sin{k_{n}d})\nonumber
\end{eqnarray}
But (\ref{fluxBull2stars}) is exactly what we would write down by combining two 3-star motifs and using (\ref{3starSecUnterm}) with $Z_n\rightarrow A_n$, $A_n\rightarrow 0$, $C_n\rightarrow B_{n}$, and an appropriate labeling of the edges.  The use of the star motif enables (\ref{fluxBull2stars}) to be written down by inspection, as the graph is a combination of two star motifs.  We immediately conclude that the secular function for the bull graph is

\begin{eqnarray}\label{secularBullgraph}
F_{bull}(a,b,c,d) &=& F_{star}(a,c,d)F_{star}(b,c,d) \\
&-& \sin k_na\sin{k_nb}\left(\sin{k_nc}+\sin{k_nd}\right)^{2}\nonumber
\end{eqnarray}

Eqn (\ref{secularBullgraph}) may be recast in a more familiar form viz.,
\begin{eqnarray}\label{secBullGraph}
F_{bull}(a,b,c,d) &=& 9\cos{k_{n}(a+b+c+d)}+\cos{k_{n}(a+b-c-d)}\nonumber \\
&-& \cos{k_{n}(a+b+c-d)}-\cos{k_{n}(a+b-c+d)}\nonumber \\
&-& \cos{k_{n}(a-b+c-d)}-\cos{k_{n}(a-b-c+d)}\nonumber \\
&-& 3\cos{k_{n}(a-b+c+d)}-3\cos{k_{n}(a-b-c-d)}\nonumber \\
&+& 16\sin{k_{n}a}\sin{k_{n}b}
\end{eqnarray}

Note that the bullgraph has no loop-only wave functions.  The amplitudes at the two internal vertices are easily calculated from (\ref{fluxBull2stars}).  The transition moments are calculated using these and the solutions to (\ref{secBullGraph}) in (\ref{xNM}).

\subsection{Diagrammatical rules for general graphs}
Every graph in table \ref{tab:resultsTable} may be solved using the fundamental star and lollipop motifs in the same way as for the bullgraph.  The details of how this is done are presented in \ref{sec:lollipopAppendix} for the lollipop-like geometries and in \ref{sec:barbellAppendix} for the barbell-like geometries.  Here, we wish to note some general rules for using motifs to solve graphs.  We will again limit the discussion to internal vertices of degree equal to three or less, but the generalization to internal vertices of arbitrary degree is straightforward and requires use of the $N-star$ motifs for degree $N$.  An example of how to do this for $N=4$ is presented in \ref{sec:starAppendix}.  We also limit the discussion to irrationally-related edges, but the generalization to rationally-related edges is detailed in \ref{sec:starAppendix} using the 3-star graph as a model.

The general method to writing down a secular equation for a graph with $N_{v}$ internal vertices is as follows:  (1) Label the vertices with their amplitudes $A,B,C,\ldots$ where flux flowing into or out of each is conserved and must flow along edges connected to the vertex and to other parts of the graph, (2) determine how the vertex and its edges overlays other vertices and their connected edges in order to identify the motifs comprising the graph, (3) use the secular functions in (\ref{3starSecUnterm}) and (\ref{lollipopSecUnterm}) with appropriately relabeled amplitudes to write a set of simultaneous equations relating the secular functions at an internal vertex to the connecting amplitudes via the motif equations, and (4) set the determinant of the amplitude matrix to zero to obtain the secular equation for the entire graph.  This process is illustrated in the Appendices for all of the graphs in table (\ref{tab:resultsTable}) as well as for more complex graphs comprised of many motifs and, in some cases, motifs that are themselves graphs that are composites of stars and lollipops.

The secular equation of a graph is generally transcendental but is easily solved using numerical methods.  Details of the statistical properties and root separators of the spectra of quantum graphs supporting a self-adjoint Hamiltonian are available in the literature, as are explicit solutions of the form $k_{n}=f(n)$ in periodic orbit expansions, but it is much simpler to solve a transcendental equation numerically for the eigenvalues for the graphs presented in this paper.

The secular equation provides a set of eigenvalues for the states of the graph where the amplitude of the particle on each edge is nonzero.  Graphs containing closed loops, such as the lollipop or barbell, can also have wavefunctions where the amplitude along a connecting or terminal edge is exactly zero, as was described for the lollipop motif in this section or for the barbell in \ref{sec:barbellAppendix}.  When multiple sets of wavefunctions are present, they must be ordered in energy and their eigenstates interleaved so that a complete set results for the graph.  Finally, graphs with no external connections, such as a barbell or triangle, will necessarily have a zero-energy eigenstate where the wavefunction over the entire graph is constant.  This ground state must be included in the spectrum in order to maintain completeness of the eigenstates.  The barbell example in the appendix demonstrates this explicitly and shows how the TRK sum rules only work when this state is included.  For most composite graphs, there will not be any additional sets of spectra other than those from the secular equation.  Again, the rational case is an exception, allowing wavefunctions that vanish at the shared vertices and form exact half-periods over each edge.  These are straightforward to handle, should they arise, and do not require solution to any transcendental secular equation.

\section{Scaling and universality of optimized composite graphs}\label{sec:results}

It is a common goal of the study of nonlinear optical structures to determine which configurations might exhibit near-maximum values.  This challenging goal has been difficult to achieve for decades; most known molecular systems fall far short of the theoretical maximum values.  Monte Carlo studies have identified states and spectra leading to large nonlinear effects, but the form of the Hamiltonian of the system remains difficult to determine.  Potential optimization studies have shown that systems comprised of piecewise continuous potentials can achieve at best $\beta_{int}\leq 0.7089\beta_{max}$.  The quantum star graphs were the first structures shown to achieve over 80 percent of the potential-optimized value \cite{lytel13.01}, and table \ref{tab:resultsTable} shows that composites of stars and loops can exceed the response of the basis star motif.  It is desirable to know whether any physical system for which a potential may be written down and solved can lead to unity.  The work in this paper has shown that quantum graphs might provide a set of deterministic models for large-scale exploration of states and spectra that can break the potential optimization barrier.

Quantum graphs are comprised of wires, loops, and stars, and the spectrum and eigenstates of any graph may be calculated using the motifs described in this paper.  The transition moments are calculable from the eigenstates, and the hyperpolarizabilities for any graph are obtained by a sum over states, as in (\ref{sh-betaMax}) and (\ref{sh-gammaMax}).  Quantum graphs are a rich toolbox for fundamental studies of the nonlinear optical limits of quantum systems because their Hamiltonians are known and solvable for any given geometry of a specific topological structure.  This means that exploration of a range of geometries of a given topology is tantamount to exploring the limits of systems with an infinite number of sets of transition moments for a fixed set of spectra.  Exploration of many topologies then provides an enormous subset of all possible sets $(x_{nm},E_{n})$ of spectra and states that satisfy the Thomas-Reiche-Kuhn sum rules, from which the fundamental limits specified in (\ref{sh-betaMax}) and (\ref{sh-gammaMax}).  Thus, the study of the behavior of quantum graphs as their geometry is varied for a given topology, and across many topologies using the motif method to calculate spectra, is a simple means to model the behavior of known physical systems as their nonlinearities approach their maximum values.

The character of quantum graphs comprised of star and lollipop motifs is dominated by the properties of the motifs.  Star and lollipop graphs have large intrinsic first and second hyperpolarizabilities, implying that composites containing stars and lollipops will have topological characteristics enabling geometric realizations with large hyperpolarizabilities.  Geometric constraints can reduce the dynamic range of the hyperpolarizability tensors by limiting the projections of the transition moments onto a specific external axis.  Further constraints, such as a closed topology with no external edges, can significantly alter the range of response for the graph.\cite{shafe12.01}

Wires, loops, and stars have spectra that are (more or less) evenly spaced.  Wires and loops have fixed energy-level spacing, whereas three-prong star graphs have fixed spacing between so-called root separators that divide the spectrum into cells of equal width, each containing a single energy level.  The variation in spectra enabled by altering the lengths of the prongs of the star graph are due precisely to the variability between levels permitted by the root boundaries, but the achievement of any desired energy level separation, such as that achieved in the Monte Carlo studies that generated near unity maxima, is not possible in a single star graph.  But as shown in this paper, many of the composite graphs in table \ref{tab:resultsTable} have nonlinearities larger than the three-prong star.  These same graphs have nonuniform root separators, and certain topological combinations of edge lengths enable variable level spacing that more closely resembles that achieved in the sum-rule-constrained Monte Carlo studies.  We anticipate that a sufficiently complex graph may be devised such that the level spacing of the most significantly-contributing levels could be near-optimum for achieving the maximum nonlinearity.  The motif method described herein enables the identification of the characteristic equation of such graphs and should prove valuable to future studies of the fundamental limits.

Additionally, classes of graphs containing star and lollipop motifs exhibit the same (universal) behavior in scaling variables as their first hyperpolarizability approaches its maximum value whenever this value is at least half the fundamental limit.  Scaling appears to result from the dynamics of the graph that causes only three levels to contribute substantially near the maximum values, the so-called three-level ansatz (TLA).  In this limit, the description of the nonlinear optics of the graphs becomes simple and universal for graphs with the star motif.  The level spacing variation enabled by composite graphs appears to approach that of the best potential-optimized models.\cite{szafr10.01,zhou06.01,zhou07.02} Both topological control and scaling universality of the hyperpolarizabilities are discussed in this section.

\subsection{Topological control of the hyperpolarizability tensors}

The global properties of the hyperpolarizability tensors are thus determined by the topology of the graphs, while the local properties, such as the projections onto a fixed external axis, are determined by the geometry of a particular realization of the graph.  For a given Monte Carlo run on a specific topology, a complete sampling of possible geometries yields the ranges of the first and second hyperpolarizability, as well as the contributions to these tensors from their spherical components.  Graphs with identical shapes but different topologies necessarily have different spectra, though the projection of their edges onto a fixed external axis could be similar.  Topological shifts alter the spectra, changing both the values and the energy-level spacing; these factors set the limits on the maximum achievable hyperpolarizability in the graph, even when the geometry is optimized for that graph.

To study a particular typological class, we sample its configuration space using Mote Carlo methods. The coordinates of the edges that define the shape are selected at random; and, the intrinsic first and second hyperpolarizabilities calculated.  The distribution of results over many configurations provides insights into the relationship between a topological class and its nonlinear-optical properties.

Figure \ref{fig:barbellTensors} illustrates the approach when applied to two distinct topologies that span the same geometries; two realizations of a barbell, one containing closed bells and the other having two open bells.  The former contains two 3-star vertices but the stars are closed into loops, and the entire graph is sealed, as explained earlier in the paper.  The open barbell graph has two open stars connected in such a way that flux travels across the structure, rather than around the loops.  The energy levels of the graphs are quite different, and so are the hyperpolarizability tensors.  Similar results hold for composite graphs comprised of the star and lollipop motifs, though the ranges of the spherical tensor components vary according to the topology.  The results in table \ref{tab:resultsTable} were obtained in this way.

\begin{figure}
\includegraphics[width=3in]{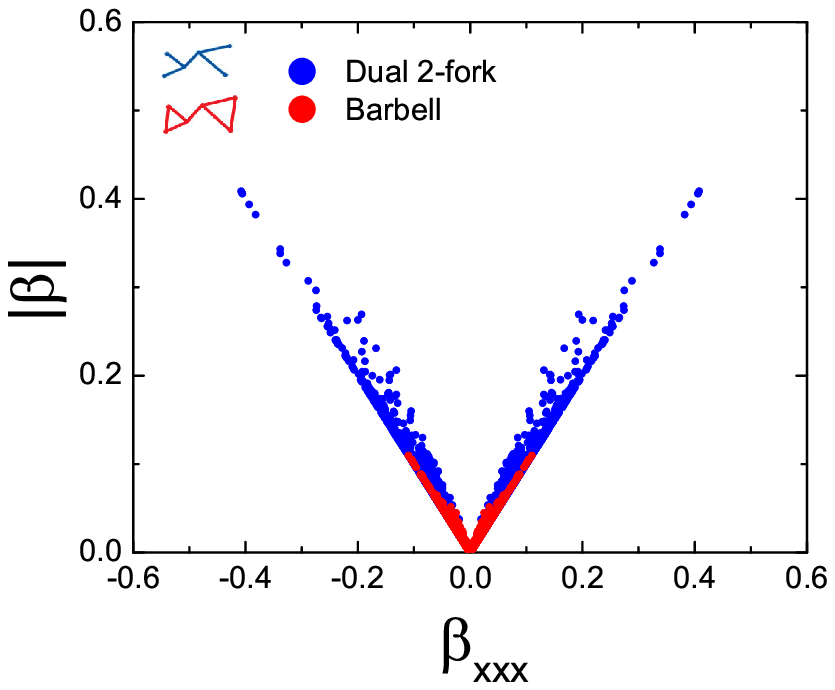}
\includegraphics[width=3in]{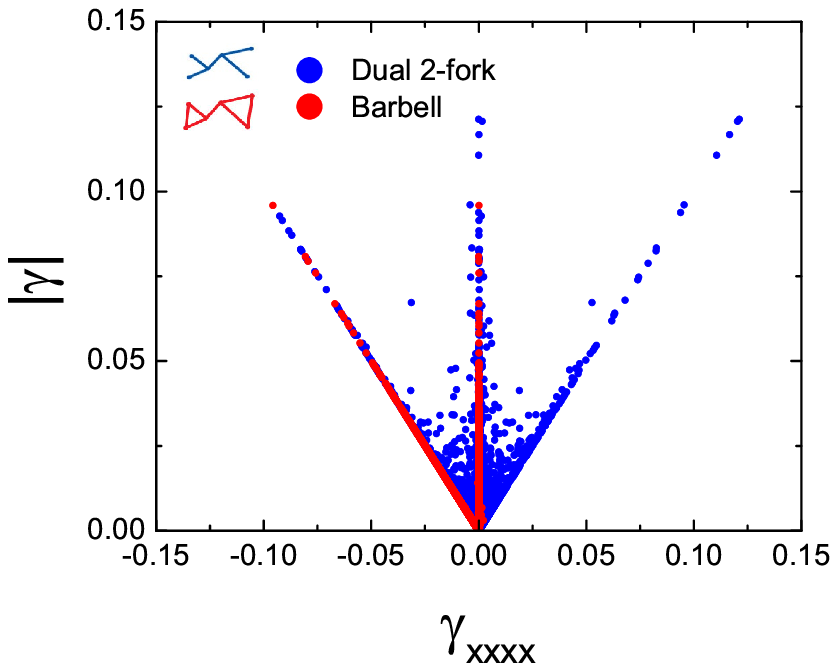}\\
\caption{Hyperpolarizability tensors and their norms for the barbell graph with two open ends (star-to-star) and two closed ends (bells) for a large sampling of shapes using a Monte Carlo method.  The profound change in the nonlinear response due to the topological change from a closed dual-loop configuration to a geometrically similar one that is isomorphic to two back-to-back 3-star graphs is self-evident.}
\label{fig:barbellTensors}
\end{figure}

Figure \ref{fig:betaGammaRanges} illustrates the change in magnitude and range offered by a simple shift in the topology of the graph for the star, lollipop, and barbell topologies.  Most significant is the observation that graphs containing star motifs with at least one prong tied to infinite potential yield consistently large intrinsic hyperpolarizabilities, even when one of the stars is closed into a loop (as in the lollipop).  But isolated, closed loop graphs always have much lower responses.  Bent wire graphs fall between these two extremes, but tend toward suboptimum response, regardless of their geometry.  This suggests that optimum configurations of one-electron, quasi-one dimensional confined systems necessarily will have a free edge and at least one star vertex.

\begin{figure}\centering
\includegraphics[width=4in]{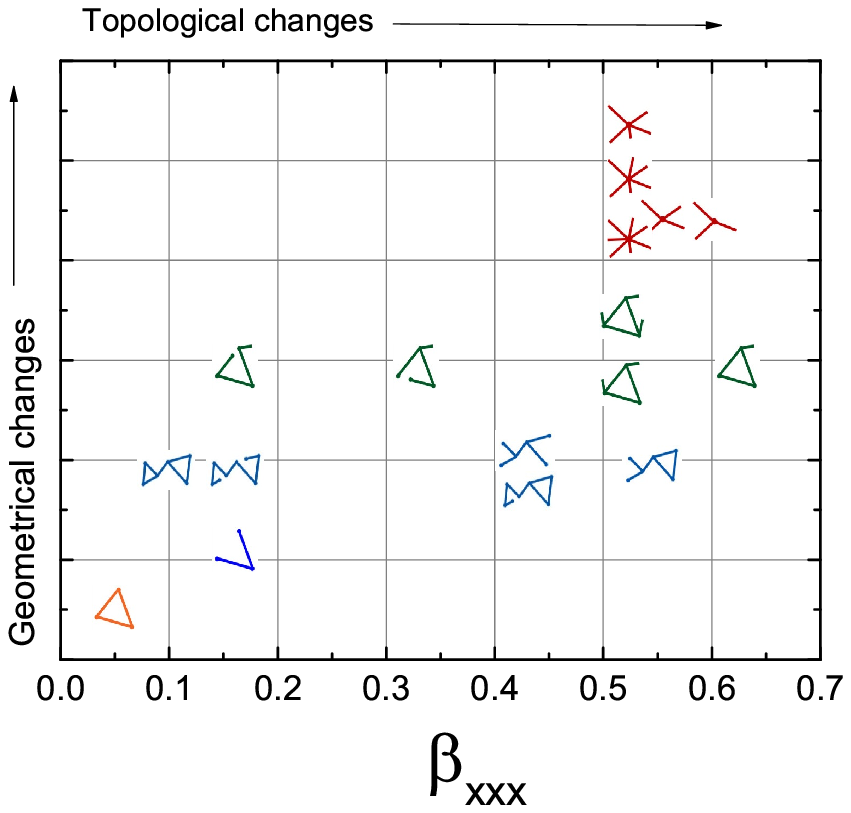}\\
\includegraphics[width=4in]{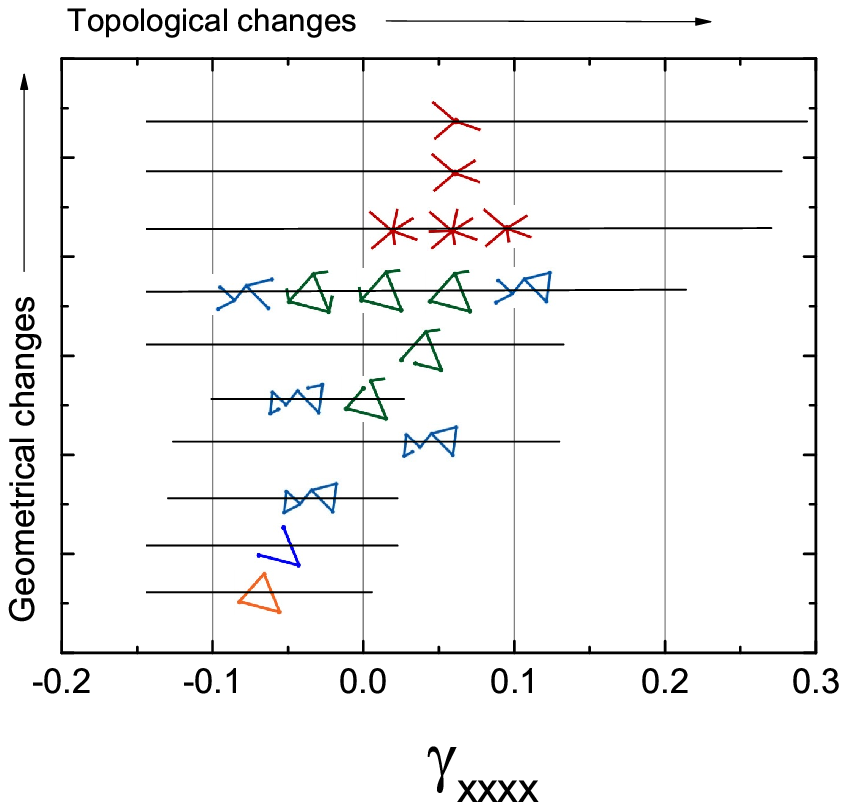}\\
\caption{Complete range of the first (top) and second (bottom) hyperpolarizability (horizonal axis) for the loop, wire, and star topologies (vertical axis) with various geometries for each. The vertical bins that change with $\beta_{xxx}$ (and $\gamma_{xxx}$) show that graphs with similar topologies have essentially the same hyperpolarizability.}
\label{fig:betaGammaRanges}
\end{figure}

Figure \ref{fig:starProngDist} illustrates the impact of edge length variarions in 3-star graphs.  The relative edge lengths set the energy spectrum of the graph and also contribute to their projection onto an external x-axis once their angular positions are specified.  For a given set of prong lengths, the value of $\beta_{xxx}$ will vary over a range as the angles between the prongs change.  However, for each set of prong lengths, there will be one set of angles for which $\beta_{xxx}$ is maximum.  The figure was constructed so that the largest values were plotted on the top. For example, stars with prongs $(1,0.6,0.13)$ appear to have the largest $\beta_{xxx}$, but this is true only if the angles take on specific values.  Underneath the contours showing the greatest values for this set of prong lengths, there are points with lesser values corresponding to the same prong lengths but nonoptimum angles; this is evident from figure \ref{fig:thirdProngVariationBeta}.  The inset in figure \ref{fig:starProngDist} shows the shapes with the largest values (red), as well as one with a much smaller value(blue).  The significance of the edges and the angles in determining which graphs have optimum geometry will be discussed in the next section.

\begin{figure}\centering
\includegraphics[width=3.4in]{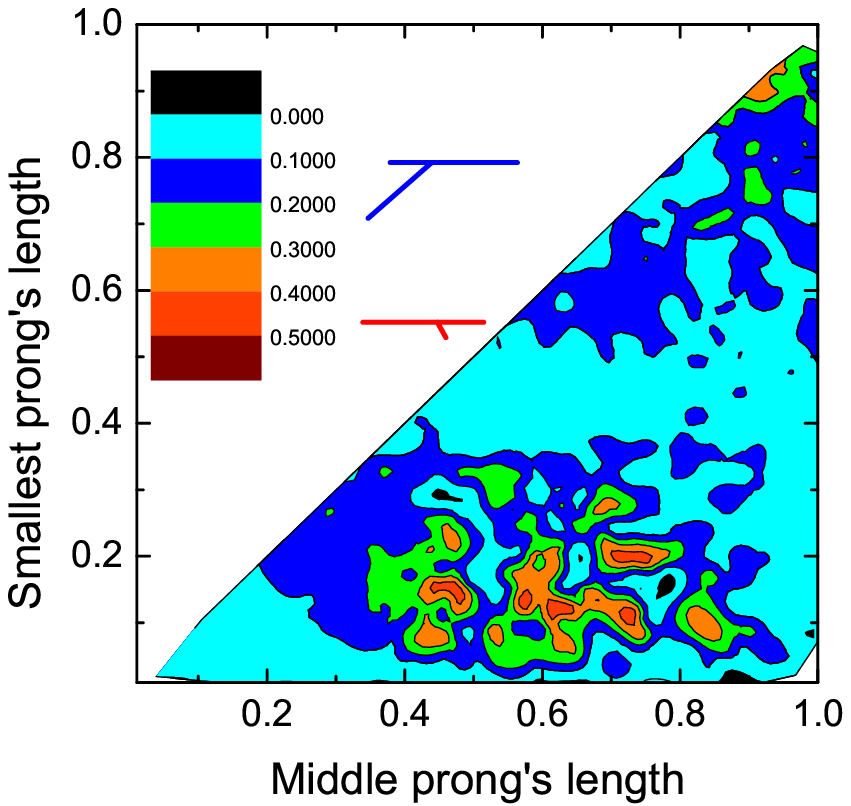}
\caption{Contour plot of the largest values of $\beta_{xxx}$ for 3-star graphs as a function of prong lengths.  The largest prong always has length unity (since the results are scale-independent).  The length of the middle prong ranges from zero to unity, while that of the shortest prong ranges from zero to a maximum equal to the middle prong. The angles each makes with the longest prong are random.  Each pair (short, middle) of prong lengths has a set of angles where $\beta_{xxx}$ is near zero, but only the optimum pairs (short, middle) can generate large $\beta_{xxx}$ for special sets of angles.  The inset shows the shape with the largest (red) and smallest (blue) $\beta_{xxx}$.}
\label{fig:starProngDist}
\end{figure}

\begin{figure}\centering
\includegraphics[width=3.4in]{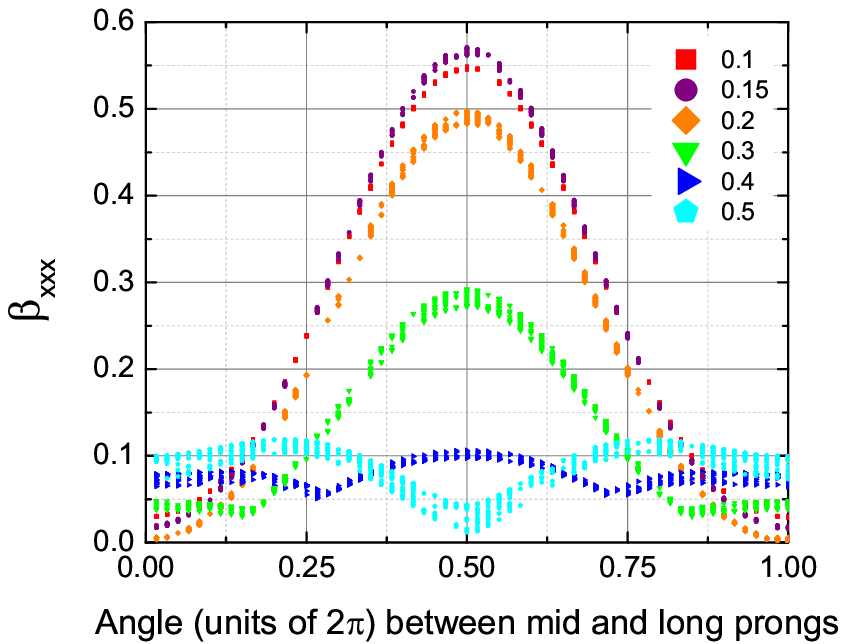}
\caption{Variation of $\beta_{xxx}$ with the angle between the middle prong of length 0.6 and the long prong of length unity for several short prong lengths.  The vertical range of points for a specific curve representing a short prong length is for the full range of angles of the small prong.  The key feature determining the strength of the nonlinearity is the antiparallel middle and large prongs, with a short prong at any angle.  A short prong permits the largest flux to move across the graph without diverting any of it into another direction.  Increasing the short prong length dramatically decreases the nonlinear response.  At a short prong length of 0.3 or greater, the nature of the angular dependence changes.}
\label{fig:thirdProngVariationBeta}
\end{figure}

The ideal star graph has its longest prong (of length one) and its second longest prong (of length $\sim 0.6$ antiparallel along $x$ and the shortest prong (of length $\sim 0.13$) at any angle.  Interestingly, a straight wire along $x$ would have zero $\beta_{xxx}$, while a bent wire could have $\beta_{xxx}\sim 0.172$ but no larger.  The attachment of a single short prong away from a center of symmetry converts the graph to a topology that generates one of the largest intrinsic values to date.

\subsection{Scaling and universality near maximum response}

The hyperpolarizabilities of nonlinear optical structures are a sum over all states of products of the transition moments divided by energies, as given in (\ref{betaInt}) and (\ref{gammaInt}).  Even when the hyperpolarizabilities are near-zero, many states contribute to the sum but their contributions tend to cancel one another.  In early work, it was predicted \cite{kuzyk00.01} that a given structure whose topology and geometry are such that it has near-optimum response will have at most three contributing states.  This three-level ansatz (TLA) appears to be a universal property of nonlinear optical systems, though there is no proof to date.

Heuristically, the TLA would appear to be a natural consequence of the result that the maximum nonlinear response is predicted by a three-level model.  Potential optimization studies \cite{zhou06.01,zhou07.02, ather12.01} support this observation but yield at best a $\beta_{xxx} \simeq 0.71$.  All calculations to date that originate with a conventional Hamiltonian appear to obey this observation, though Monte Carlo studies using randomly selected states and transition moments consistent with the TRK sum rules can exceed this limit and approach unity \cite{kuzyk08.01}.  The disparity is the subject of ongoing research and will not be further discussed here.  Returning to the potential optimization models, the structures having the maximum nonlinear response universally have the property that $X = x_{01}/x_{01}^{max} \simeq 0.79$, where $x_{01}^{max}$ is given by
\begin{equation}\label{Xmax}
x_{01}^{max} = \left(\frac{\hbar^2}{2 m E_{10}}\right)^{1/2} ,
\end{equation}
with $E_{10}=E_{1}-E_{0}$.  In the three-level ansatz (TLA), the normalized first hyperpolarizability $\beta_{xxx}$ can be expressed as \cite{kuzyk09.01}
\begin{equation}\label{betaIntfG}
\beta_{xxx} = f(E)G(X) ,
\end{equation}
where
\begin{equation}\label{f(E)}
f(E) = (1-E)^{3/2} \left( E^2 + \frac {3} {2} E + 1 \right),
\end{equation}
and
\begin{equation}\label{G(X)}
G(X) = \sqrt[4]{3} X \sqrt{\frac {3} {2} \left( 1 - X^4\right)},
\end{equation}
and $E=E_{10}/E_{20}$.

\begin{figure}\centering
\includegraphics[width=3in]{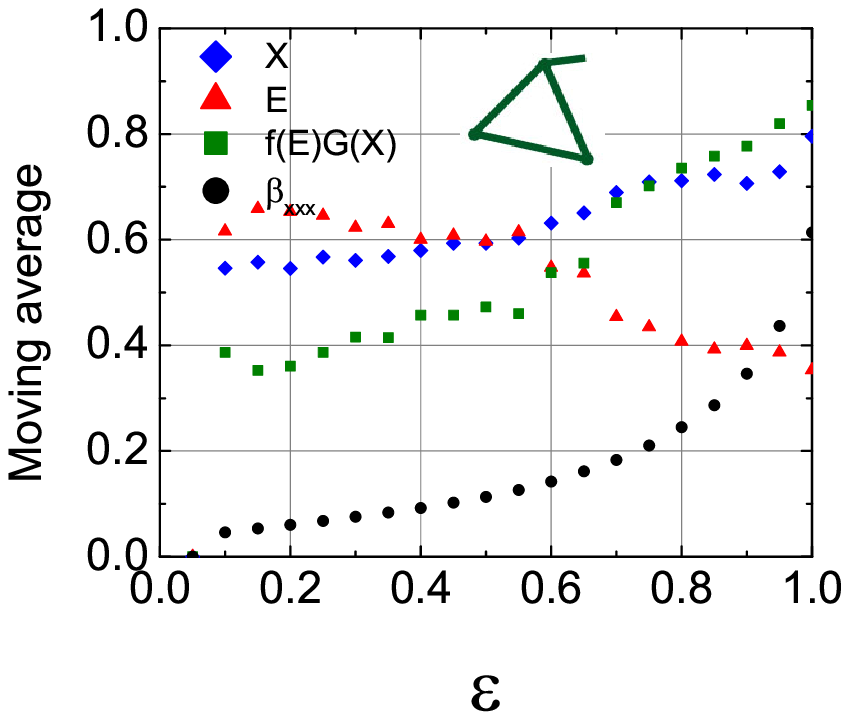}\includegraphics[width=3in]{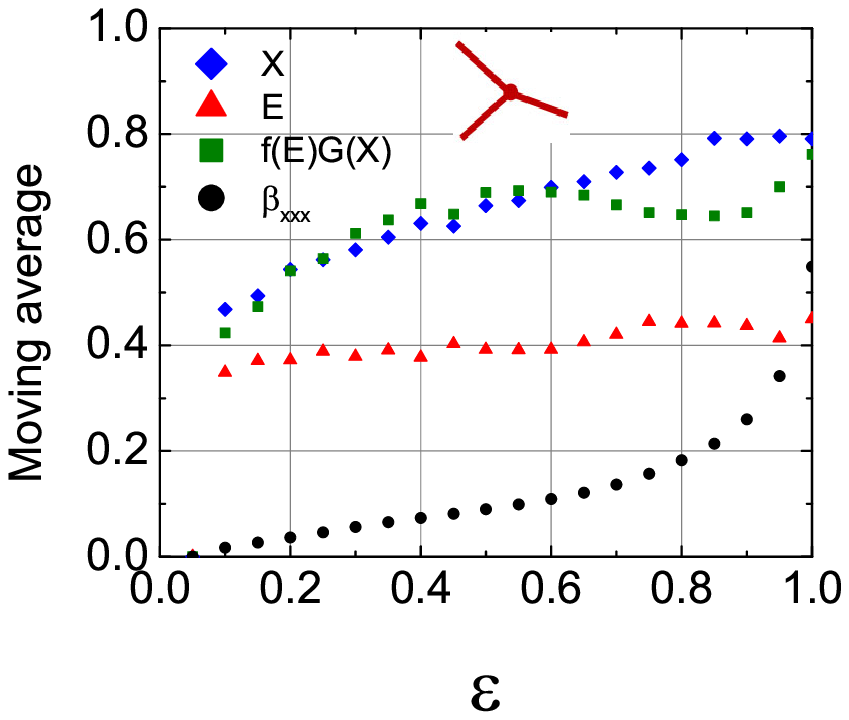}\\
\includegraphics[width=3in]{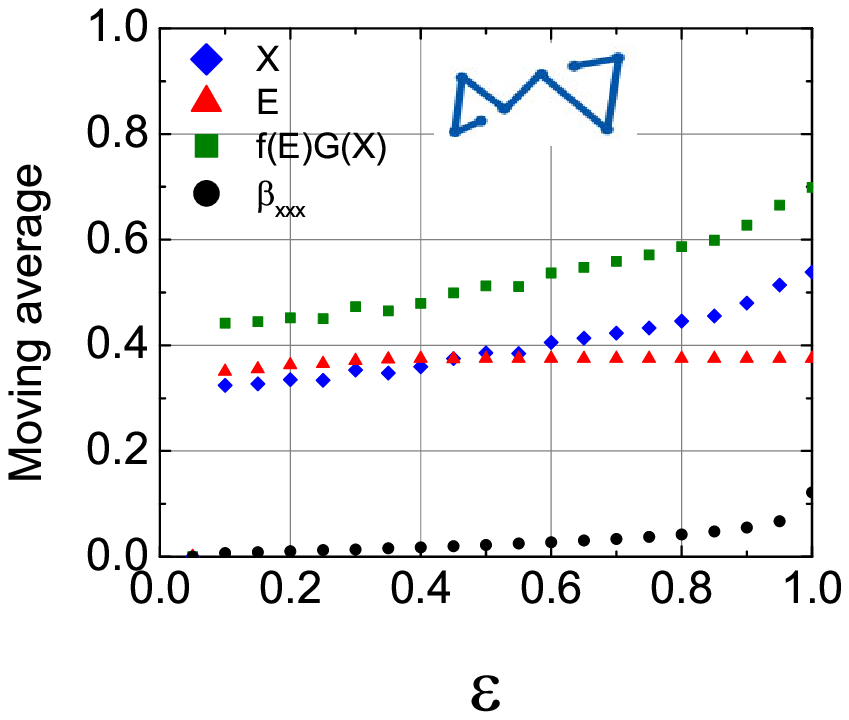}\includegraphics[width=3in]{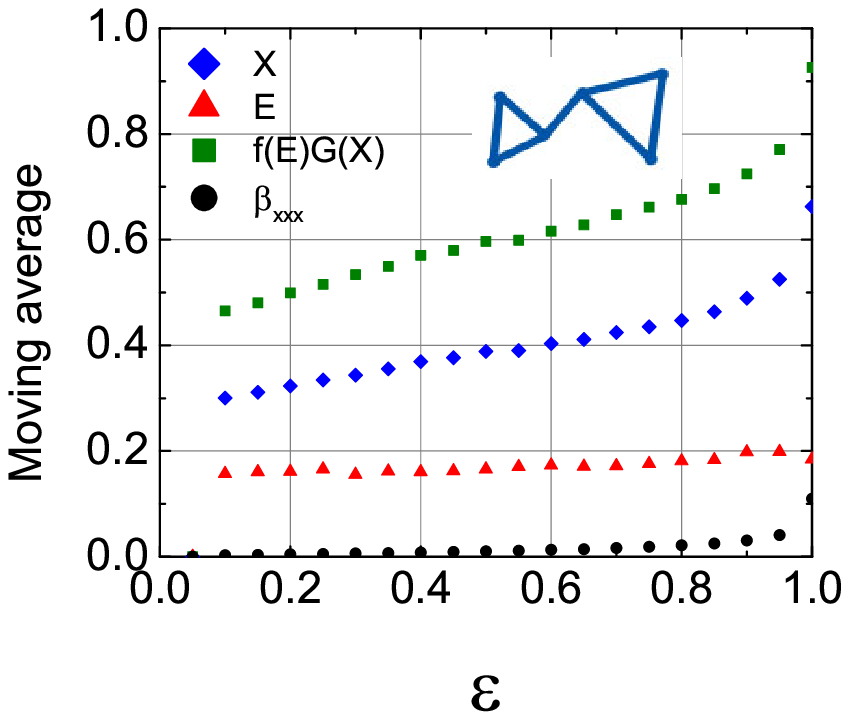}\\
\caption{Scaling of $X$, $E$, and the three-level Ansatz value of $\beta_{xxx}\equiv f(E)G(X)$ for the best topologies as their optimum geometry is approached.  The parameter $\epsilon$ is the ratio of $\beta_{xxx}$ for a particular geometry to the value of the best geometry.  As the best configurations for the star and lollipop are approached, $X$, $E$, and $fG$ scale toward the universal values because these topologies are near-optimum.  For the geometrically similar but topologically distinct barbell and seven-wire (barbell opened up), the response is far from optimum, and the values of $E$, $X$, and $fG$ do not approach those of the optimum three-level model; instead there is a gap between the best $\beta_{xxx}$ and $f(E)G(X)$.}
\label{fig:XEfG}
\end{figure}

This expression is valid as the hyperpolarizability approaches its maximum intrinsic value.  It might be anticipated that quantum graphs containing star motifs meeting the criteria of having at least one open prong and at least one star vertex would have spectra and states producing the universal values of E and X as their geometry approaches the optimum shape. In other words, we anticipate that near-optimum graphs with star motifs would have a first hyperpolarizability whose value approaches that predicted by the optimum three-level model, $f(E)G(X)$. Figure \ref{fig:XEfG} shows that this is a valid conclusion for both the star and lollipop, each of which has $\beta_{xxx}>0.55$, nearly 80 percent of the theoretical maximum for systems that derive from a Hamiltonian.\cite{zhou06.01,zhou07.02,szafr10.01}  Equally profound is the converse observation that the geometrically similar but topologically-distinct bent seven-wire and closed barbell, whose topologies are far from optimum, show a large gap between their maximum $\beta_{xxx}$ and $f(E)G(X)$, with values of E and X not equal to the universal values.

In figure \ref{fig:starProngDist}, we noted that the star with the optimum set of prong ratios had the largest response for a specific set of angles, and that stars with other sets of prong ratios could not be altered by changing their angles to achieve a response as large.  We also noted that if the angles of the star with the optimum edge ratios were changed from optimum, the response would drop, and one would see sample points behind those indicating the optimum values in the figure with the same prong ratios but different angles.  This is corroborated by two additional observations.  First, the star graph in figure \ref{fig:XEfG} has a nearly constant value of E throughout the range of responses, indicating that the changes in the spectra are not the dominant driver behind the optimization, but instead the angles and transition moments (as indicated by the value of X in the figure) drive the optimum configuration.  An inspection of the lowest spectra of the star graph confirms this.

In figure \ref{fig:starSpectrumBeta}, the ten lowest momentum eigenvalues are displayed for a Monte Carlo run where the ensemble members are ordered such that their maximum $\beta_{xxx}$ increases from left to right.  The numerical value of $\beta_{xxx}$ is displayed as the dashed curve and its value shown on the right axis in figure \ref{fig:starSpectrumBeta}.  The eigenvalues each vary in a random way between their root separators as the geometry of the star is altered from left to right until the optimum geometries are attained.  When the maximum is approached, the lowest eigenvalues converge to well-defined values, as reflected by the plots of $X$, $E$, and $fG$ in the figure \ref{fig:XEfG} for star graphs.

\begin{figure}\centering
\includegraphics[width=4in]{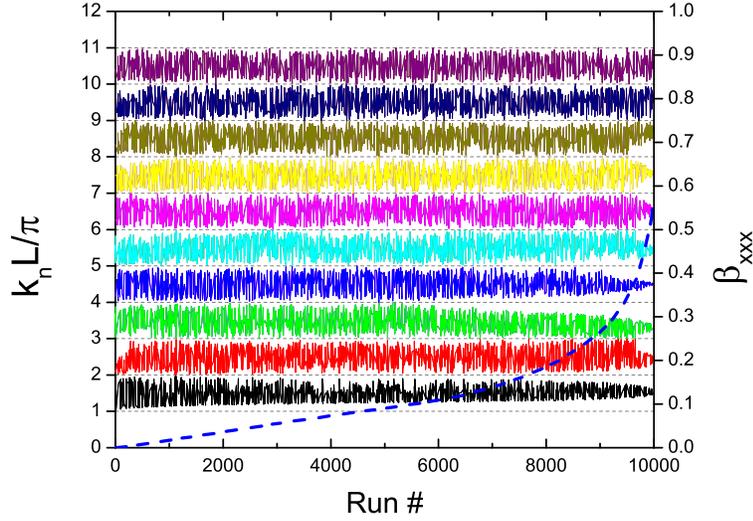}
\caption{Variation of the lowest ten momentum eigenvalues of the 3-star graph for $10,000$ randomly generated samples, ordered by their $\beta_{xxx}$ values.  For the 3-star, the solutions to the secular equation lie between fixed root boundaries located at multiples of $\pi/L$, with L equal to the sum of all edges.  The three lowest momentum eigenvectors asymptote to fixed values as $\beta_{xxx}$ (shown as a dashed curve) approaches its maximum value for the best geometry.}
\label{fig:starSpectrumBeta}
\end{figure}

This observation is not true for the lollipop graph, whose spectrum is a complex interleaving of two sets of disparate spectra (one each for the entire graph and for the loop-only part), as demonstrated earlier in the paper and illustrated in figure \ref{fig:lollipopSpectrumAndBeta}.  For these graphs, both E and X change dramatically as the geometry is changed toward the optimum shape, as indicated in figure \ref{fig:XEfG}.  Figure \ref{fig:closeupLollipopSpectrum} is a close-up of these first eight eigenstates of the lollipop spectrum to illustrate that there are always well-defined boundaries between a given set of eigenstates for a fixed run, and that somewhere between runs  3000 and 4000, $\beta_{xxx}$ begins to climb, and the modes jump to a different-looking pattern where the variation of the three lowest eigenvalues decreases rapidly and then converge to fixed values at the maximum hyperpolarizability, with a universal value of the energy ratio E.

Since the maximum value of $\beta_{xxx}$ for lollipops is larger than that of the basic star graph, we might expect that further changes in the complexity of the spectrum of a graph could lead to even larger responses, perhaps approaching or even exceeding the potential limit of about 0.71.  In complex graphs, the root boundaries may acquire an almost random structure to them, suggesting they might be {\em tunable} to provide the kinds of level spacing required to achieve maximum nonlinear responses.  Future work will explore this possibility.

\begin{figure}\centering
\includegraphics[width=3.4in]{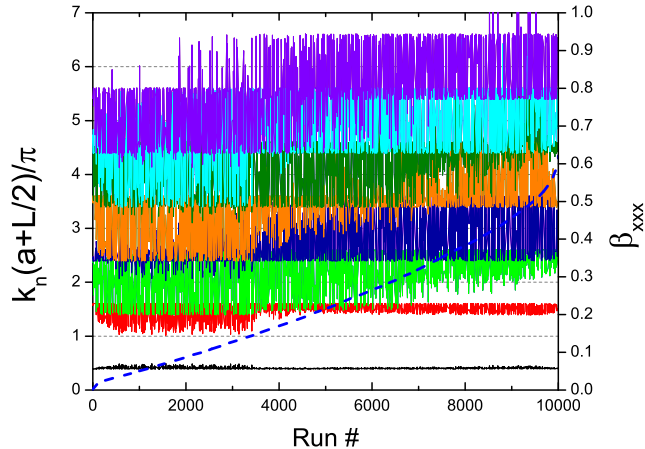}
\caption{Variation of the lowest eight momentum eignevalues of the lollipop graph for $10,000$ randomly generated samples, ordered by their $\beta_{xxx}$ values. $\beta_{xxx}$ is shown as the dashed curve.}
\label{fig:lollipopSpectrumAndBeta}
\end{figure}

\begin{figure}\centering
\includegraphics[width=3in]{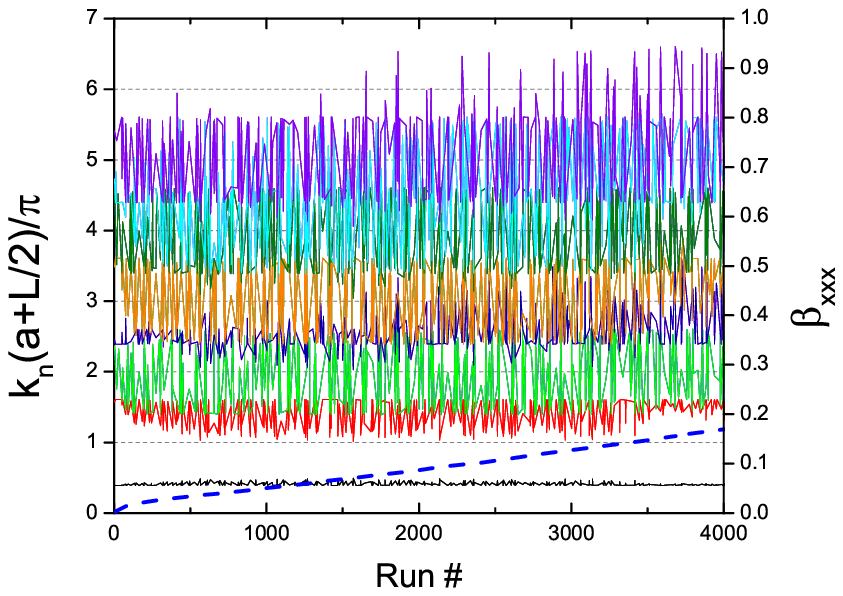}\includegraphics[width=3in]{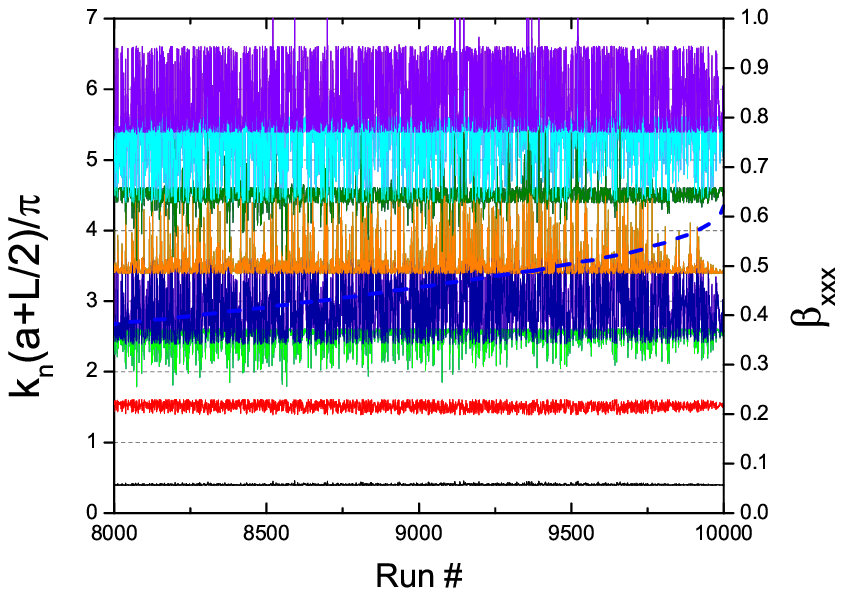}\\
\includegraphics[width=3in]{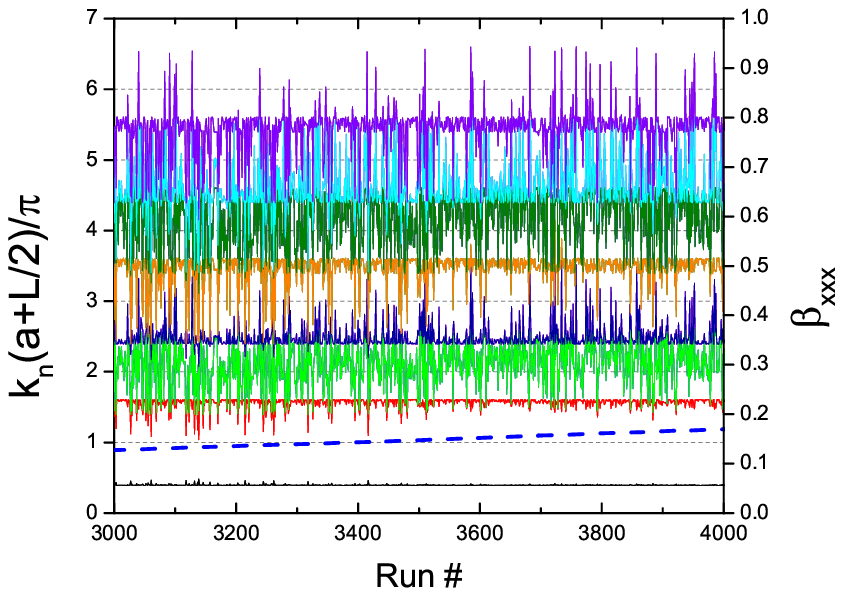}\includegraphics[width=3in]{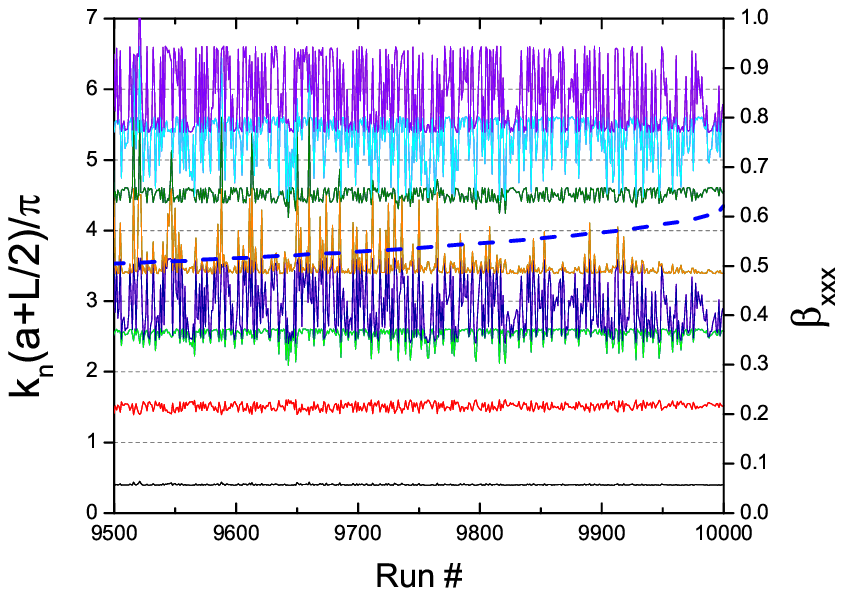}\\\caption{Close-up view of the eigenvalues for the lollipop graph in figure \ref{fig:lollipopSpectrumAndBeta}.  The dashed line shows $\beta_{xxx}$.  The top left panel shows large variation in the eigenvalues when $\beta_{xxx}$ is near zero, and reveals an upward shift of the eigenvalues as $\beta_{xxx}$ exceeds 0.1.  This region is magnified in the bottom left panel for clarity.  The top right panel shows how the first three modes converge to fixed values near maximum $\beta_{xxx}$.  The bottom right panel is a close-up view of this region.  Despite the appearance of mode overlap in the top panels, the bottom panels show clearly that the modes never overlap, but instead vary such that their ordering is maintained.}
\label{fig:closeupLollipopSpectrum}
\end{figure}

\section{Conclusions}\label{sec:outlook}

Star graphs are fundamental building blocks, or motifs, from which complex graphs may be constructed. We have presented a new method for evaluating the nonlinear optical properties of complex quantum graphs based upon the use of the star graph motif and its secular equation.  We showed how the motif method may be scaled to solve graphs comprised of many star vertices and larger motifs comprised of graphs based upon the elementary star motif.  We invoked the method to solve several geometrically similar but topologically distinct variants of the lollipop and barbell graphs, respectively.  We provided a set of rules for calculating any graph, taking into account both degeneracies from rationally-related edges as well as the appearance of multiple sets of eigenstates arising from subgraphs.  We also related the global properties of closed graphs to the appearance of a zero energy, constant amplitude ground state.

The general methods for solving for the hyperpolarizabilities of quantum graphs that we previously developed could then be used with the states and energies from the motif analysis to calculate the cartesian and spherical tensor components of the first ($\beta$) and second ($\gamma$) hyperpolarizability to understand the impact of topology across geometrically equivalent graphs on the nonlinear optical tensors.  In particular, graphs with identical topologies have comparable maximum nonlinearities, while graphs with identical geometries but different topologies have far different maximum nonlinearities.  This behavior has been previously observed for bent wires and loops \cite{shafe12.01} so it is not surprising that it holds for star graphs and their extensions.  But with the advent of the star motif for constructing the spectral equations for complex graphs, we now have a fundamental explanation for both the similar, topological responses and the differences when topologies are altered so that the underlying secular spectral functions of geometrically similar graphs no longer resemble one another.  Scaling according to the theory of fundamental limits also holds across different star geometries, so long as the star motif is active within the graph so that its global properties are dominated by the star topology.  Interestingly, the addition of a star vertex to a loop creates the lollipop graph which has one of the largest intrinsic first hyperpolarizabilities of all graphs, despite the fact that the loop by itself has a nonlinearity that is over ten times smaller.  The star vertex is key to the synthesis of molecular systems modeled by the elementary quantum graph, as it appears to guarantee that a geometrically-unconstrained star topology will have a large, intrinsic first and second hyperpolarizability.

The one-electron elemental graph model is a simple but effective way to explore a wide range of states and transition moments enabled by a structure's Hamiltonian and boundary conditions, from the bottom up, i.e., by solving the equations of motion to determine the maximum hyperpolarizabilities of a topological class of graphs for comparison with the abstract theory of fundamental limits based upon the use of the Thomas-Reiche-Kuhn sum rules in a sum over states expansion of $\beta$ and $\gamma$.

We presented an analysis of the scaling properties of the tensors of graphs as they approached their optimum geometries for maximum response.  We verified that the three-level Ansatz appears to hold for quantum graphs whose maxima are at least 80 percent of the potential optimized fundamental limit.  We were unable to find any graphs for which the hyperpolarizabilities exceeded the potential optimization limits (eg, $\beta\sim 0.71$).  The gap between this limit and the unit maximum predicted by using the TRK sum rules directly, without reference to any specific Hamiltonian remains. It is possible that more complex graphical models, with multiple electrons and dressed edges can have the richer spectra required to achieve the optimum values of the dimensionless X and E parameters that can generate responses closer to the unit fundamental limit.  However, we are leaning toward the point of view that the actual limit of the hyperpolarizability in a real nonlinear optical material will be that obtained from potential optimization and not the generalized result from invoking only the sum rules.

\ack SS and MGK thank the National Science Foundation (ECCS-1128076) for generously supporting this work.

\appendix
\section{Spectral functions for lollipop-like graphs.}\label{sec:lollipopAppendix}

Any elementary QG may be solved for its states and spectra by writing the edge functions in the canonical form (\ref{edgeFunctionAB}), conserving flux at each vertex, and solving the secular equation of the resulting amplitude equations for the eigenvalues.  But using motifs, the secular equation for a graph may almost be written down by inspection.  The key is to note that the amplitudes of the motifs in figure \ref{fig:motifGraphs} must match at vertices where the motifs are combined.  This ensures all of the flux exiting a motif through its connecting edge to another motif will enter that motif, and vice versa.  We showed in Section \ref{sec:motif} how this may be achieved for the bullgraph in table \ref{tab:resultsTable}.  Here and in the rest of the appendices, we apply the method to the remaining graphs in the table as well as several more complex graphs that would be otherwise very messy to solve.

\subsection{Lollipop bull}

Consider first a more challenging example of using the motifs to calculate the secular equation of a graph.  Figure \ref{fig:lollipopBull} shows how three star motifs are combined to produce a graph we call a lollipop-bull with a prong at each corner of a three-sided loop.  There are three vertices $D,E,F$ with nonzero amplitudes where flux is conserved.  The appropriate star motif is the one illustrated in figure \ref{fig:motifGraphs} with only one terminated end, with $A\longrightarrow D$, $B\longrightarrow E$, and $C\longrightarrow F$.  This leads to three equations for $D,E,F$ as follows:

\begin{eqnarray}\label{starAtDEF}
D_{n}F_{star}(a,d,f) &=& E_n\sin k_{n}a\sin k_{n}f + F_n\sin k_{n}a\sin k_{n}d \nonumber \\
E_{n}F_{star}(b,d,e) &=& D_n\sin k_{n}b\sin k_{n}e + F_n\sin k_{n}b\sin k_{n}d \nonumber \\
F_{n}F_{star}(c,e,f) &=& D_n\sin k_{n}c\sin k_{n}e + E_n\sin k_{n}c\sin k_{n}f \nonumber \\
&&
\end{eqnarray}

Setting the determinant of the matrix of coefficients in the set (\ref{starAtDEF}) yields the exact secular equation for the graph:
\begin{eqnarray}\label{secularPopBull}
F_{popbull} &=& G_{1}G_{2}G_{3}-G_{1}G_{1c}-G_{2}G_{2c}-G_{3}G_{3c}-G_{0},\nonumber \\
G_{1} &=& F_{star}(a,d,f)\nonumber \\
G_{2} &=& F_{star}(b,d,e) \\
G_{3} &=& F_{star}(c,e,f)\nonumber \\
G_{1c} &=& \sin{k_nb}\sin{k_nc}\sin{k_nd}\sin{k_{n}f}\nonumber \\
G_{2c} &=& \sin{k_na}\sin{k_nc}\sin{k_nd}\sin{k_{n}e}\nonumber \\
G_{3c} &=& \sin{k_na}\sin{k_nb}\sin{k_ne}\sin{k_{n}f}\nonumber \\
G_{0} &=& \sin{k_na}\sin{k_nb}\sin{k_nc}\sin{k_nd}\sin{k_ne}\sin{k_{n}f}\nonumber
\end{eqnarray}
The eigenvalues may be found by setting $F_{popbull}=0$, as usual.  The amplitude ratios may be obtained from (\ref{starAtDEF}) and are given by
\begin{eqnarray}\label{ampRatios}
E_{n}/D_{n} &=& \frac{\sin{k_{n}b}\left[G_{1}+\sin{k_{n}a}\sin{k_{n}e}\right]}{\sin{k_{n}a}\left[G_{2}+\sin{k_{n}b}\sin{k_{n}f}\right]}\nonumber \\
&& \\
F_{n}/D_{n} &=& \frac{\sin{k_{n}c}\left[G_{1}+\sin{k_{n}a}\sin{k_{n}e}\right]}{\sin{k_{n}a}\left[G_{3}+\sin{k_{n}c}\sin{k_{n}d}\right]}\nonumber
\end{eqnarray}
By using motifs, we are able to get to the secular equation for a complicated graph almost by inspection.

The matrix of coefficients for (\ref{starAtDEF}) will prove useful later when considering composite graphs with the lollipop bullgraph as a motif itself.  We write it in the reduced form
\begin{equation}\label{secMatrixPopBull}
M_{popbull}=
\begin{bmatrix}
f_{star}(a,d,f) & -\csc{k_{n}d} & -\csc{k_{n}f} \\
-\csc{k_{n}d} & f_{star}(b,e,f) & -\csc{k_{n}e} \\
-\csc{k_{n}f} & -\csc{k_{n}e} & f_{star}(c,e,f) \\
\end{bmatrix}
\end{equation}
where the reduced form of the star secular function is given in (\ref{reducedSecStar}) with appropriate edge labeling.  The secular function for the lollipop may then be expressed in reduced form $F_{popbull}=\det{M_{popbull}}$ which will be useful later.

\begin{figure}\centering
\includegraphics[width=2in]{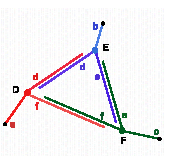}\includegraphics[width=2in]{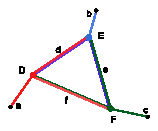}\\
\caption{The bullgraph with an extra prong is a lollipop-bullgraph that is topologically comprised of three 3-star motifs.}
\label{fig:lollipopBull}
\end{figure}

\subsection{Open lollipop (4-wire)}

Opening up the central vertex of the lollipop to release one edge turns the lollipop into a bent 4-wire graph (figure \ref{fig:OpenLollipops}, right).  The eigenmodes and energies of this graph are identical to that of a wire, but there are in general four different projections to account for.

\subsection{Open lollipop (bent wire star)}
Opening up a vertex away from the central vertex turns the lollipop graph into a 3-star graph (figure \ref{fig:OpenLollipops},left) but with one of the prongs bent into a 2-wire geometry.  The spectra of this graph are identical to those of a 3-star graph, but the transition moments take a slightly different form since one prong has a different projection onto the x-axis unless the two bent wires become parallel.

\begin{figure}\centering
\includegraphics[width=2in]{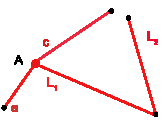}\includegraphics[width=2in]{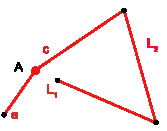}\\
\caption{Opening up a lollipop graph can produce either a 3-star or 4-wire graph with the same geometry.}
\label{fig:OpenLollipops}
\end{figure}

\section{Spectral functions for the barbell-like graphs.}\label{sec:barbellAppendix}

The barbell is shown in table \ref{tab:resultsTable}.  It is a fully-sealed graph with no prong at infinite potential.  This requires special treatment, as shown next.

\subsection{Barbell graph}

Consider a barbell graph, with two $N=3$ loops connected by a single prong, as shown in Fig \ref{fig:barbellSepCom}.  With the definitions $L_1=b_1+b_2+b_3$ and $L_2=c_1+c_2+c_3$, the eigenfunctions may be written as a union $\cup_{k=1}^3 \phi_{n}^{k}$ of edge functions for the two loops and their connector. There are three distinct sets of eigenstates for the barbell graph, one each for a set of wave functions where the particle is on one loop or the other, with zero amplitude on the connector, and a third set where the particle may be anywhere on the graph.  To see this, just note that the barbell is a composite of two lollipop motifs.  The eigenstates for the lollipop were a pair of sets, each corresponding to what would be the even and odd wave functions of an open loop graph (ie, a bent 3-wire).  The even wave functions have opposite slopes at the star vertex, so the flux on the prong may be nonzero.  The odd wave functions have identical slopes at the vertex, so the prong amplitude and flux can vanish, yielding loop-only wave functions for the graph.  For the barbell, the same situation applies, except there are two loops.

\begin{figure}\centering
\includegraphics[width=2in]{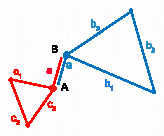}\includegraphics[width=2in]{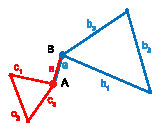}\\
\caption{Barbell graph comprised of two lollipop motif graphs.}
\label{fig:barbellSepCom}
\end{figure}

The loop-only wave functions are easy to write down, as they are simply sine functions with wavevectors $k_{n}=2\pi n/L_{m}$, where $m=1,2$ and n is a positive integer.  These sets of wave functions must be combined with the remaining set and ordered in energy in order to create a complete, ordered set of eigenfunctions for the graph.

Consider the solutions with nonzero amplitude on all edges.  From figure \ref{fig:barbellSepCom}, we can immediately write down the boundary-matching conditions that generate the secular function for the barbell.
\begin{eqnarray}\label{lollipopSecUnterm}
B_{n}F_{pop}(a,L_{2}) = A_{n}\cos k_{n}L_{1}/2 \\
A_{n}F_{pop}(a,L_{1}) = B_{n}\cos k_{n}L_{2}/2\nonumber
\end{eqnarray}
from which we conclude that
\begin{eqnarray}\label{secularBarbell2}
F_{barbell} &=& F_{pop}(a,L_1)F_{pop}(a,L_2) \\
&-&\cos k_{n}L_{1}/2\cos k_{n}L_{2}/2\nonumber
\end{eqnarray}
It is straightforward to show (after some algebra) that Eqn (\ref{secularBarbell}) is identical to the following form:
\begin{eqnarray}\label{secularBarbell}
F_{barbell} &=& 9\sin\left(k_n(a+(L_1+L_2)/2)\right)\nonumber \\
&+&\ \ \sin\left(k_n(a-(L_1+L_2)/2)\right) \\
&-& 3\sin\left(k_n(a+(L_1-L_2)/2)\right)\nonumber \\
&-& 3\sin\left(k_n(a-(L_1-L_2)/2)\right)\nonumber
\end{eqnarray}
Knowing the motifs enabled us to get $F_{barbell}$ in just a few steps.

The ground state of this graph has $k=0$, ie, zero energy but with a nonzero amplitude.  This state corresponds to a particle equally likely to be anywhere on the graph, and has zero flux transmitted anywhere in the graph.  The reason this state exists in this graph but not in the prong-loop, star, or bent wire graphs (despite the fact that all of these have nondegenerate wave functions) is the same reason it exists in the single closed loop (triangle, quad, quint) graphs (which do have degenerate wave functions) and has nothing to do with degeneracy:  It exists because there is no $\emph{anchor-to-zero}$ edge attached to the graph.  Such an edge has a terminal vertex with a Dirichlet boundary condition requiring the wavefunction to vanish there, so a constant solution would necessarily have to take the value zero.  For a closed network with no such terminal vertices, the constant in the network may be nonzero.  This means the ground state will always be a zero-energy state for these type of graphs.

Fig \ref{fig:barbellAllandNoZeroSum} displays the results of calculating the sum rules for the barbell graph, but this time, we use all of the states including (left) and excluding (right) the zero energy ground state.  The exquisite sensitivity of the TRK sum rules reveals the existence of the zero energy ground state and verifies that this state and the three sets of eigenmodes are required to get a complete set of eigenstates for this graph.

\begin{figure}\centering
  \includegraphics[width=80mm]{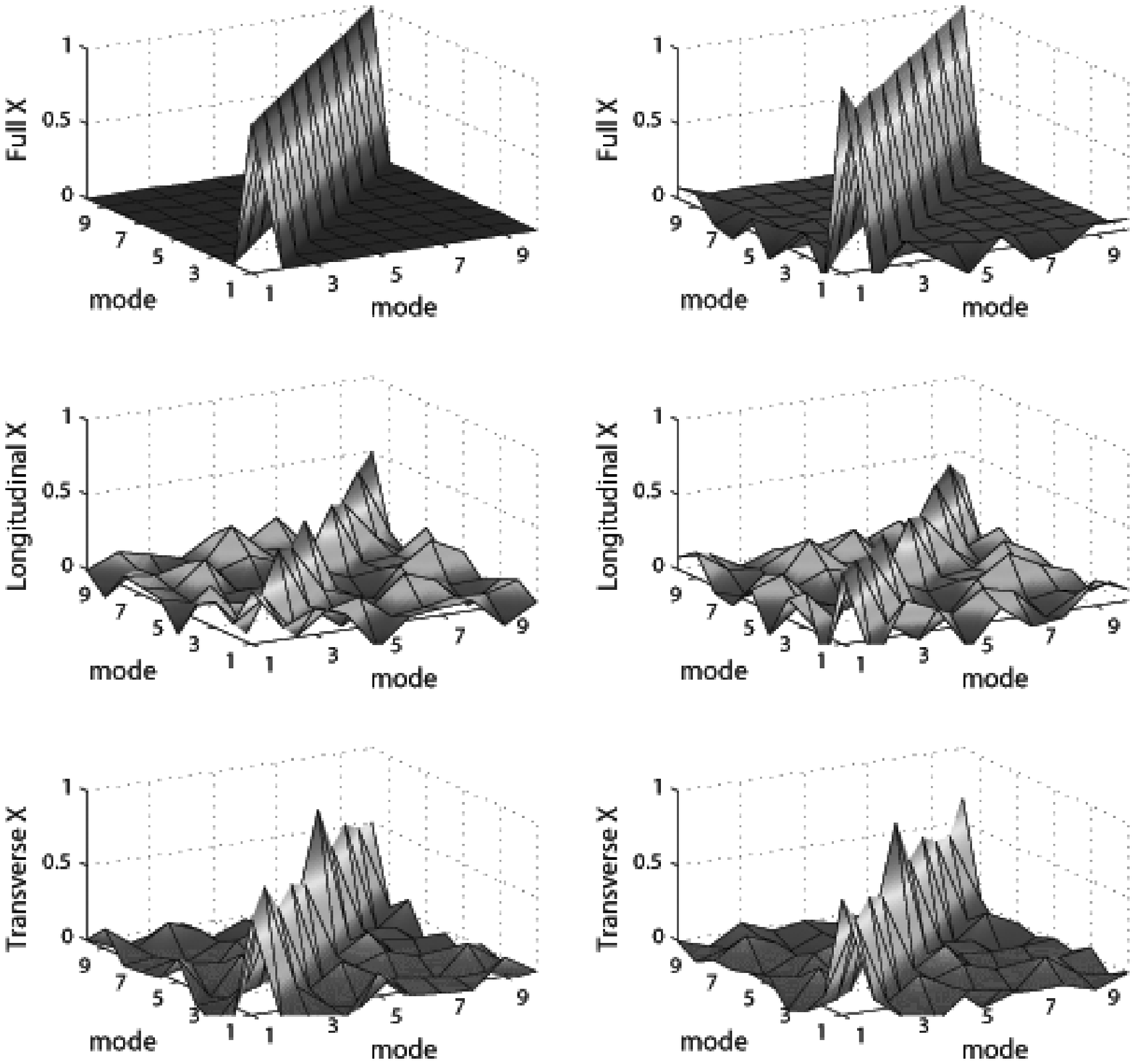}\\
  \caption{Sum rules for the set of states including the zero energy ground state (left column) and excluding it (right column).  The top row is the sum of the longitudinal sum rule (middle row) and the transverse sum rule (bottom row).  The exquisite sensitivity of the sum rules indicates the states are only complete if the zero energy ground state is included for the barbell graph.  The X in the vertical axis labels indicates that these are the sum rules calculated using the $x_{nm}$ transition moments.}\label{fig:barbellAllandNoZeroSum}
\end{figure}

\subsection{Open barbell (star-stick lollipop)}

We are now in a position to use the secular functions for the star and lollipop motifs to solve for the secular equation of the combined graph in Fig \ref{fig:lollipop2starstick}.  We see from the Figure that we should connect the lollipop to the star such that vertex A of the star is vertex Z of the lollipop, and vice-versa.  We also set $B_{n}=C_{n}=0$ to terminate the other two prongs of the star. This results in the relationship
\begin{eqnarray}\label{lollipop2starAmps}
A_{n}F_{star}(a,b,c) &=& B_{n}\sin k_{n}b\sin k_{n}c \\
B_{n}F_{pop}(a,L_{tot}) &=& A_{n}\cos k_{n}L_{tot}/2 \nonumber
\end{eqnarray}
Cross-multiplying (or setting the determinant of the coefficients to zero) yields the secular function $F_{star-pop}(a,b,c,d,e)$ for the eigenvalues of the star-stick lollipop graph:
\begin{eqnarray}\label{seclollipop2star}
F_{star-pop}(a,b,c,d,e) &=& F_{star}(a,b,c)F_{pop}(a,L_{tot}) \\
&-& \sin k_{n}b\sin k_{n}c\cos k_{n}L_{tot}/2 \nonumber
\end{eqnarray}
The solutions to $F_{star-pop}(a,b,c,d,e)=0$ are the eigenvalues of the star-stick lollipop graph.  The amplitudes $A_{n}$ and $B_{n}$ are then found from Eqn (\ref{lollipop2starAmps}). With these in hand, the hyperpolarizabilities for this class of graphs are calculated as described previously in Section \ref{sec:QGreview}.

\begin{figure}\centering
\includegraphics[width=2in]{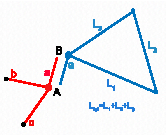}\includegraphics[width=2in]{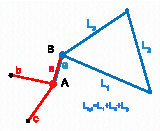}\\
\caption{The hybrid star-lollipop graph formed from the union of the star and lollipop motifs.}
\label{fig:lollipop2starstick}
\end{figure}

\subsection{Open barbell (dual 2-fork)}

Consider the graph in Fig \ref{fig:star2star} with two star vertices connected by a common prong.  This graph is obtained from the prior one by breaking one of the edges in the loop. There are two central vertices connected by an edge, and each is a 3-star motif with two ends at zero amplitude.  The coupled amplitude equations are easy to write down using Eqn (\ref{3starSecUnterm}) with $B_n=C_n=0$. They are
\begin{eqnarray}\label{star2starAmps}
A_{n}F_{star}(a,b,e)=B_{n}\sin k_na\sin k_nb\nonumber \\
B_{n}F_{star}(c,d,e)=A_{n}\sin k_nc\sin k_nd
\end{eqnarray}
The secular function for this graph is thus
\begin{eqnarray}\label{secularStar2star}
F_{star-star} &=& F_{star}(a,b,e)F_{star}(c,d,e) \\
&-& \sin{k_na}\sin{k_nb}\sin{k_nc}\sin{k_nd}\nonumber
\end{eqnarray}

\begin{figure}\centering
\includegraphics[width=1.7in]{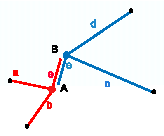}\includegraphics[width=1.7in]{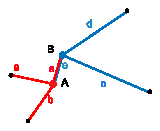}\\
\caption{The hybrid star-to-star graph formed from the union of two star motifs.}
\label{fig:star2star}
\end{figure}

The secular function in Eqn (\ref{secularStar2star}) may be rewritten in the following form:
\begin{eqnarray}\label{SecularFcnS2S2}
F_{sec} &=& -4\sin{k_{n}a}\sin{k_{n}b}\sin{k_{n}c}\sin{k_{n}d}\sin{k_{n}e}\nonumber \\
&-& 2\sin{k_{n}(a+b+c+d+e)}\nonumber \\
&+& \ \sin{k_{n}(a+b-c-d+e)}\nonumber \\
&-& \ \sin{k_{n}(a+b-c-d-e)}\nonumber \\
&+& .5\sin{k_{n}(a+b+c-d+e)}\nonumber \\
&+& .5\sin{k_{n}(a+b+c-d-e)}\nonumber \\
&+& .5\sin{k_{n}(a+b-c+d+e)} \\
&+& .5\sin{k_{n}(a+b-c+d-e)}\nonumber \\
&+& .5\sin{k_{n}(a-b+c+d+e)}\nonumber \\
&+& .5\sin{k_{n}(a-b+c+d-e)}\nonumber \\
&-& .5\sin{k_{n}(a-b-c-d+e)}\nonumber \\
&-& .5\sin{k_{n}(a-b-c-d-e)}\nonumber
\end{eqnarray}

\subsection{Open barbell (bent stick lollipop)}

If we disconnect one edge in one of the barbell loops, we will get a 4 wire lollipop, Fig \ref{fig:OpenBarbells}, left, from which we conclude that the secular function is identical to that of a lollipop but with the prong length equal to $a+c_1+c_2+c_3\equiv a+L_2$, with $L_1=b_1+b_2+b_3$ and $L_2=c_1+c_2+c_3$.  The amplitude equations are identical to those of the lollipop with these substitutions, and the secular function becomes
\begin{eqnarray}\label{secular4wirePop}
F_{4wire pop} &=& \frac{1}{2}\left[3\cos\left(k_n(a+L_{2}+L_{1}/2)\right)\right.\nonumber \\
&-& \left.\cos\left(k_n(a+L_{2}-L_{1}/2)\right)\right]
\end{eqnarray}
Setting $F_{4wire pop}=0$ yields the wavenumbers for the eigenfunctions where the particle may be found on any of the edges in the barbell.  Since this is a lollipop graph, there are also another set of eigenstates for motion on the loop only.

\begin{figure}\centering
\includegraphics[width=1.7in]{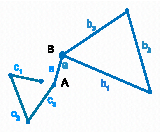}\includegraphics[width=1.7in]{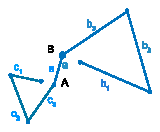}\\
\caption{Opening up a barbell graph can produce either a bent stick lollipop or a 7-wire graph with the same geometry.}
\label{fig:OpenBarbells}
\end{figure}

Note that a loop may also be opened at the vertex connecting $c_1$ with $c_2$ or $c_2$ with $c_3$, and the resulting graph would become a star-stick lollipop as in Fig \ref{fig:lollipop2starstick}, but with one of the star prongs bent so the graph is geometrically a barbell.  The secular equation will be identical to that of the star-stick lollipop.

\subsection{Open barbell (7-wire)}

Figure \ref{fig:OpenBarbells},right, also shows a barbell graph where the loops are opened at their central vertices.  This converts the topology to that of a 7-wire graph, bent into the barbell shape.  The calculation of this graph is trivial using the sequential path method \cite{lytel12.01} with the results shown in Table \ref{tab:resultsTable}.

\section{Motifs to calculate more complex graphs}\label{complexAppendix}

The power of motifs is easily extended to graphs with greater complexity than any in table \ref{tab:resultsTable}.  Consider the graph in figure \ref{fig:snowflakeStar}, comprised of three of the star motifs attached to a central star motif.  A straightforward calculation of the entire set of matched amplitudes and fluxes would produce a set of nine coupled equations for the nine edges.  But using motifs, there are only four relevant equations to consider:

\begin{eqnarray}\label{snowflakeStarAmps}
A_{n}F_{star}(a_1,a_2,a_3) &=& Z_n\sin k_{n}a_2\sin k_{n}a_3 \nonumber \\
B_{n}F_{star}(b_1,b_2,b_3) &=& Z_n\sin k_{n}b_2\sin k_{n}b_3 \nonumber \\
C_{n}F_{star}(c_1,c_2,c_3) &=& Z_n\sin k_{n}c_2\sin k_{n}c_3 \nonumber \\
Z_{n}F_{star}(a_1,b_1,c_1) &=& A_n\sin k_{n}b_1\sin k_{n}c_1 \\
&+& B_n\sin k_{n}a_1\sin k_{n}c_1 \nonumber \\
&+& C_n\sin k_{n}a_1\sin k_{n}b_1 \nonumber
\end{eqnarray}

Setting the determinant of the matrix of coefficients in the set (\ref{snowflakeStarAmps}) yields a compact and exact secular equation for the graph:

\begin{eqnarray}\label{snowFlStSec}
&&F_{star}(a_1,a_2,a_3)F_{star}(b_1,b_2,b_3) \nonumber \\
&\times&F_{star}(c_1,c_2,c_3)F_{star}(a_1,b_1,c_1) \nonumber \\
&=& F_{star}(a_1,a_2,a_3)F_{star}(b_1,b_2,b_3) \nonumber \\
&\times&\sin k_{n}a_1\sin k_{n}b_1\sin k_{n}c_2\sin k_{n}c_3 \nonumber \\
&+& F_{star}(a_1,a_2,a_3)F_{star}(c_1,c_2,c_3) \\
&\times&\sin k_{n}a_1\sin k_{n}c_1\sin k_{n}b_2\sin k_{n}b_3 \nonumber \\
&+& F_{star}(b_1,b_2,b_3)F_{star}(c_1,c_2,c_3) \nonumber \\
&\times& \sin k_{n}b_1\sin k_{n}c_1\sin k_{n}a_2\sin k_{n}a_3 \nonumber
\end{eqnarray}
The solutions to (\ref{snowFlStSec}) may be obtained numerically by intersection the two sides of the equation but will not be further discussed here.

\begin{figure}\centering
\includegraphics[width=1.7in]{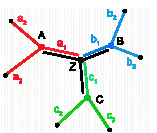}
\includegraphics[width=1.7in]{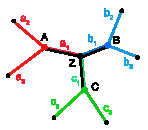}\\
\caption{A complex graph comprised of four 3-star motifs.}
\label{fig:snowflakeStar}
\end{figure}

The graph of figure \ref{fig:lollipopBull}, a three-prong composite of three 3-star motifs, is itself insertable into the graph of Figure \ref{fig:snowflakeStar} in place of the star, as shown in figure \ref{fig:snowflakeStarLollipop}.  The graph then becomes a composite of six, 3-star motifs, or of three, 3-star motifs and one lollipop bullgraph.  It is useful to demonstrate how to incorporate composites as motifs into even more complex graphs.  We will use unrationalized secular functions for the star (\ref{reducedSecStar}) and the lollipop bull (\ref{secMatrixPopBull}) in what follows.  The amplitude equations describing the movement of flux around the graph may be written in matrix form.  With the definitions
\begin{equation}
V_{star}=
\begin{bmatrix}
A_{n} \\
B_{n} \\
C_{n} \\
\end{bmatrix}
,\ V_{popbull}=
\begin{bmatrix}
D_{n} \\
E_{n} \\
F_{n} \\
\end{bmatrix}
\end{equation}
\begin{equation}
M_{1}=
\begin{bmatrix}
\csc{k_{n}a_{1}} & 0 & 0 \\
0 & \csc{k_{n}b_{1}} & 0 \\
0 & 0 & \csc{k_{n}c_{1}} \\
\end{bmatrix}
\end{equation}
\begin{equation}
N=
\begin{bmatrix}
f_{star}(a_1,a_2,a_3) & 0 & 0 \\
0 & f_{star}(b_1,b_2,b_3) & 0 \\
0 & 0 & f_{star}(c_1,c_2,c_3) \\
\end{bmatrix}
\end{equation}
and the definition of the secular function for the lollipop bullgraph in (\ref{secMatrixPopBull}), we may write the amplitude equations as
\begin{eqnarray}\label{snowPopBullAmps}
N V_{star} &=& M_{1} V_{popbull} \\
M_{popbull} V_{popbull} &=& M_{1} V_{star} \nonumber
\end{eqnarray}

These are almost in the canonical form for motifs, since $N$ is diagonal and its elements are the secular functions for the stars at their central vertices.  Using the relationship $M^{-1}\det(M)\equiv M^{adj}$ between a matrix M and its adjoint $M^{adj}$, with $M\equiv M_{popbull}$, we arrive at
\begin{equation}\label{snowPopBull2ndAmps}
F_{popbull} V_{popbull}=M_{popbull}^{adj}M_{1} V_{star}
\end{equation}
which puts the amplitude relationship in canonical form for a motif.  In general, the secular function of a graph will multiply the vector of its amplitudes when it is used as a motif inside a bigger graph.
For this graph, the secular function is
\begin{equation}\label{secSnowPopBull}
F_{snowpopbull}=\det{\left(M_{popbull}-M_{1}N^{-1}M_{1}\right)}
\end{equation}
which is a function of the twelve edge lengths in the graph.  Note that $\det (M_{popbull})$ is the secular function of the lollipop bullgraph, so (\ref{secSnowPopBull}) expresses the secular function of the composite of the three stars and the lollipop bull in terms of the secular functions of the individual motifs.

\begin{figure}\centering
\includegraphics[width=1.7in]{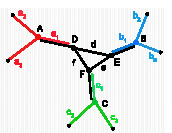}\includegraphics[width=1.7in]{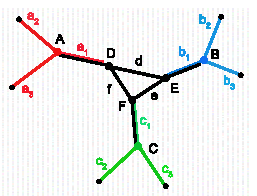}\\
\caption{A complex graph comprised of three 3-star motifs attached to a lollipop-bullgraph used as a central motif.}
\label{fig:snowflakeStarLollipop}
\end{figure}

The simplification of computations provided by the use of motifs is just one of their features.  Future work will address the self-similar scaling of nonlinear optical graphs, which may be described by using motifs that are repeatedly combined or inserted into existing graphs, but with a scale change at each insertion step.

\section{Spectral functions for star graphs}\label{sec:starAppendix}

We have previously described how to solve 3-stars.  Here, we show how the results generalize for stars with $N\geq 4$.

\subsection{Scaling to $N\geq 4$ star vertices}

Referring to figure \ref{fig:4starMotif}, we easily generalize (\ref{3starEdges}) to arrive at the 4-edge equivalent of (\ref{3starSecUnterm}) and get
\begin{eqnarray}\label{4starSecUnterm}
Z_{n}F_{4star}(a,b,c,d) &=& A_n\sin{k_{n}b}\sin{k_{n}c}\sin{k_{n}d} \nonumber \\
&+& B_n\sin{k_{n}a}\sin{k_{n}c}\sin{k_{n}d} \\
&+& C_n\sin{k_{n}a}\sin{k_{n}b}\sin{k_{n}d}\nonumber \\
&+& D_n\sin{k_{n}a}\sin{k_{n}b}\sin{k_{n}c}\nonumber
\end{eqnarray}
where the secular function for the 4-star graph is given by the four-edge version of (\ref{reducedSecStar}) times $\prod_{j=1}^{4}\sin{k_{n}a_{i}}$ with appropriate relabeling of the edges.  After some simple algebra, the secular function may be written as
\begin{eqnarray}\label{4starSecularF}
F_{4star}(a,b,c,d) &=& \frac{1}{2}\left[\sin{k_{n}(a+b)}\cos{k_{n}(c-d)}\right. \nonumber \\
&+& \left. \cos{k_{n}(a-b)}\sin{k_{n}(c+d)}\right.\nonumber \\
&-& \left. \sin{k_{n}(a+b+c+d}\right].
\end{eqnarray}
This form may be used as a motif to solve composite graphs with several four-edge vertices, such as the bubble graph shown in figure \ref{fig:4starMotif}, for which the secular function may again be written down by inspection:
\begin{eqnarray}\label{4starBubbleSecularF}
&& F_{4bubble}(a,b,c,d,e_1,e_2,f_1,f_2)= \\
&& F_{4star}(a,b,L_1,L_2)F_{4star}(c,d,L_1,L_2)\nonumber \\
&-& \sin{k_{n}a}\sin{k_{n}b}\sin{k_{n}c}\sin{k_{n}d}(\sin{k_{n}L_{1}}+\sin{k_{n}L_{2}})^{2}\nonumber
\end{eqnarray}
where $L_{1}=e_{1}+e_{2}$, $L_{2}=f_{1}+f_{2}$ are the two (sequential) bubble edges.
\begin{figure}\centering
\includegraphics[width=1.7in]{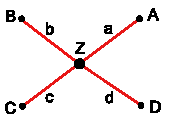}\includegraphics[width=3.4in]{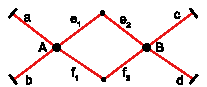}\\
\caption{4-star motif with four open edges carrying flux to/from another motif to which their edges might be attached.}
\label{fig:4starMotif}
\end{figure}
Generalizing to any number of edges is straightforward and will not be further discussed in this paper.

\subsection{Handling degeneracies}\label{sec:degeneracies}

Throughout this paper, the edges of the stars have been constrained to be irrationally-related.  This ensures the canonical form of the edge functions in (\ref{edgeFunctionAB}) may be used without reservation, as the edge functions never vanish at the central vertices.  We briefly examine rationally-related edges to show how to solve these, too.
The isolated star graph can have doubly-degenerate states for wavenumbers satisfying $k_{n}=n \pi/L$, with $L=a+b+c$ for certain values of the edges. If we write the edge functions for the three prongs as
\begin{eqnarray}\label{edgeStates3starDeg}
\phi_{n}^{(1)}(s_{1}) &=& A_{n}\sin{k_{n}(a-s_{1})}\nonumber \\
\phi_{n}^{(2)}(s_{2}) &=& B_{n}\sin{k_{n}(b-s_{2})} \\
\phi_{n}^{(3)}(s_{3}) &=& C_{n}\sin{k_{n}(c-s_{3})}\nonumber
\end{eqnarray}
then the amplitudes at the center satisfy
\begin{equation}\label{3starAmps}
A_{n}\sin{k_{n}a} = B_{n}\sin{k_{n}b} = C_{n}\sin{k_{n}c},
\end{equation}
and conservation of flux yields the secular function for an isolated star graph as
\begin{equation}\label{secDeg}
F_{star} = A_{n}\cos{k_{n}a} + B_{n}\cos{k_{n}b} + C_{n}\cos{k_{n}c}.
\end{equation}
If none of the sine functions in (\ref{3starAmps}) vanishes, then the wavenumbers satisfy the usual secular equation, (\ref{3starSecularF}).  The derivative of the secular function is
\begin{equation}\label{secDerivDeg}
-dF_{star}/dk = aA_{n}\sin{k_{n}a} + bB_{n}\sin{k_{n}b} + cC_{n}\sin{k_{n}c}.
\end{equation}
For irrationally-related edges, $F_{star}(k_{n})=0$ determines the nondegenerate eigenvalues $k_{n}$, and the derivative $dF_{star}/dk$ is never zero for $k=k_{n}$.  But when the edges are rationally-related, both the secular equation and its derivative will occasionally vanish for the same $k$, the doubly-degenerate eigenvalues.  When this occurs, (\ref{secDeg}) and (\ref{secDerivDeg}) may be used to extract amplitudes for a pair of orthogonal, degenerate states corresponding to the same eigenvalue, because the same secular equation holds for the degenerate case, as well \cite{pasto09.01}, as can be shown through a scattering matrix solution or simply by noting that the transition from irrationally-related edges to rationally-related ones is equivalent to an infinitesimal change in the arguments of (\ref{secDeg}).  Consequently, one may move from one case to the other by performing all the divisions used to derive (\ref{3starSecularF}) and then taking the rational limit.
To see how this comes about, examine the spectrum in figure \ref{fig:secPlots}.  The vertical lines at $k_{n}=n\pi/L$ (with $L=a+b+c$) are the root separators, defining cells in which only one root may be found.  For certain values of the edges, there are roots on either side of a root separator that converge toward each other and meet at a separator (becoming degenerate roots) as the edge values are tweaked toward specific ratios.  When this happens, all three terms in (\ref{3starAmps}) vanish, and they also vanish in the derivative of the secular equation (which is why the roots are doubly-degenerate).  A single degenerate root has a pair of eigenstates whose amplitudes are determined by the secular equation (\ref{secDeg}) and the requirement that the pair of degenerate states are orthogonal.  If the edge coefficients are labeled $(A_{1}B_{1}C_{1})$ and $(A_{2}B_{2}C_{2})$, the orthogonality condition is $aA_{1}A_{2}+bB_{1}B_{2}+cC_{1}C_{2}=0$.  A suitable set of coefficients may then be determined from this and the secular relations, with $A_1=1$, $A_2=1$, $C_1=1$ as the roots converge to $k_{r}$ as follows:
\begin{eqnarray}\label{degAmpCoeff}
B_{1} &=& -\frac{\cos{k_{r}c}+\cos{k_{r}a}}{\cos{k_{r}b}}\nonumber \\
C_{2} &=& -\frac{a\cos{k_{r}b}-bB_{1}\cos{k_{r}a}}{c\cos{k_{r}b}-bB_{1}\cos{k_{r}c}}\nonumber \\
B_{2} &=& -\frac{\cos{k_{r}a}+C_{2}\cos{k_{r}c}}{\cos{k_{r}b}}
\end{eqnarray}
where the cosines will take the values $\pm 1$ as their arguments each approach their own multiple of $\pi$.
\begin{figure}
\includegraphics[width=3in]{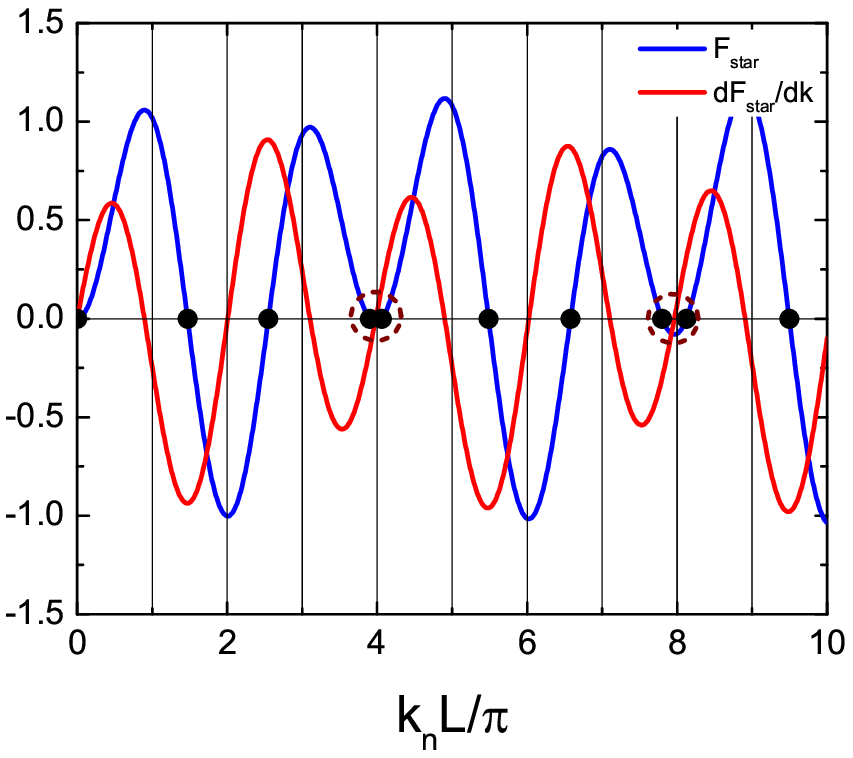}\includegraphics[width=3in]{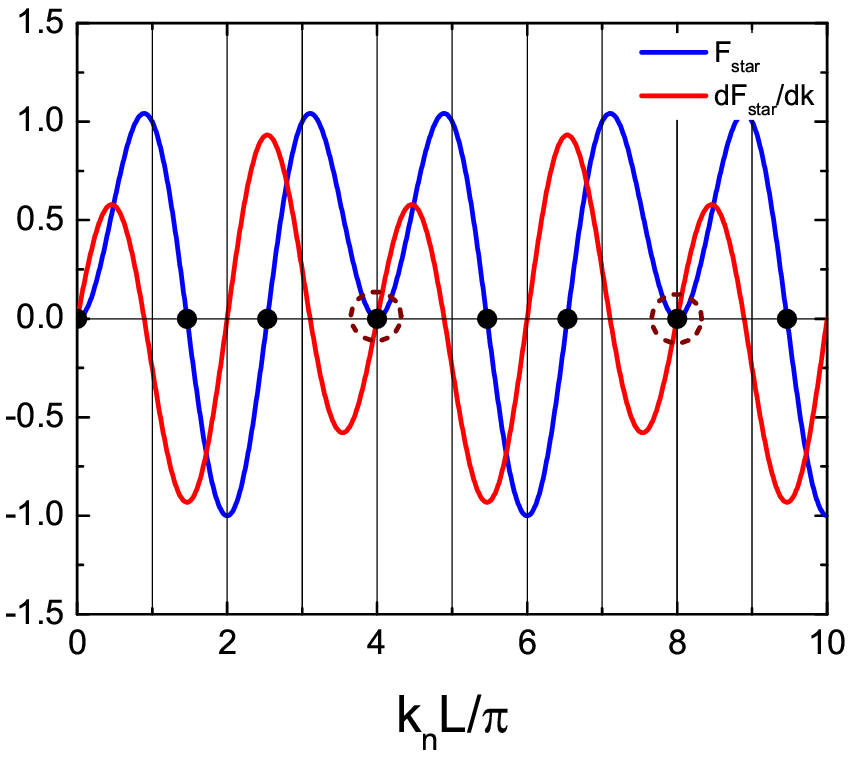}\\
\caption{Solutions to the secular equation of a 3-star graph with prong lengths that are irrationally-related (top) but approach a rational relationship (bottom).  The two roots enclosed by the dotted circles in the irrational case (one each on either side of a root separator) coalesce into a pair of degenerate roots (on a root separator) enclosed by the dotted circle, as the lengths become rationally-related  The transition is smooth in that there is no abrupt change in the nonlinear response of the graph as the edges become rationally-related.}\label{fig:secPlots}
\end{figure}
Even when there are no degeneracies for a given set of rationally-related edges, it is possible that one or two of the sine functions in (\ref{3starAmps}) could vanish.  (If all three vanish, then the root is a degenerate root boundary).  In this case, the amplitudes may still be obtained by using the amplitude (\ref{3starAmps}) and the secular equation.  For example, suppose a given solution to the secular equation $k_{m}$ satisfies $\sin{k_{m}a}=0$, but that $\sin{k_{m}b}\neq 0$ and $\sin{k_{m}c}\neq 0$.  Then (\ref{3starAmps}) and the secular equation yield the singlet solution set
\begin{eqnarray}\label{oneSineZero}
B_{m} &=& C_{m}\frac{\sin{k_{m}c}}{\sin{k_{m}b}} \\
A_{m} &=& -B_{m}\frac{\sin{k_{m}(b+c)}}{\cos{k_{m}a}\sin{k_{m}c}}\nonumber \\
\end{eqnarray}
The single unknown coefficient $C_{m}$ is determined by normalization, of course.  When two sine functions vanish, say $\sin{k_{m}a}=0$ and $\sin{k_{m}b}=0$, the solutions are even easier to obtain.  Then (\ref{3starAmps}) yields $C_{m}=0$ and $A_{m}=-B_{m}\cos{k_{m}b}/\cos{k_{m}a}$.  This solves the degenerate case for any relationship among the edges.  The correctness of the solutions may be verified by using the double commutator form of the TRK sum rules for quantum graphs \cite{shafe12.01}.  Figure \ref{fig:degSumRuleP} displays the correct (left) and incorrect (right) results when the degenerate states are included (left) or excluded (right) from the eigenstates and spectrum.
\begin{figure}\centering
\includegraphics[width=80mm]{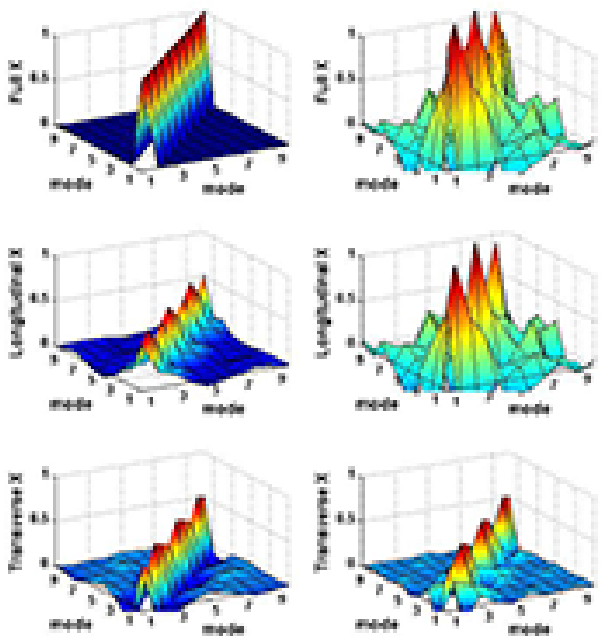}
\caption{Total (top), longitudinal (middle), and transverse (bottom) sum rules when including (left) and ignoring (right) degenerate states.}\label{fig:degSumRuleP}
\end{figure}
The extension of this analysis to graphs comprised of star motifs is straightforward, but it provides no additional information to that obtained from the nondegenerate case.


\begin{thebibliography}{10}

\bibitem{maker65.02}
P.~D. Maker and R.~W. Terhune.
\newblock {Study of Optical Effects Due to an Induced Polarization Third-Order
  in the Electric Field Strength}.
\newblock {\em Phys. Rev.}, 137(3A):801--818, 1965.

\bibitem{bloem68.01}
N.~Bloembergen, R.~K. Chang, S.~S. Jha, and C.~H. Lee.
\newblock {Optical Second-Harmonic Generation in Reflection from Media with
  Inversion Symmetry}.
\newblock {\em Phys. Rev.}, 174(3):813--22, 1968.

\bibitem{bass69.01}
M.~Bass, D.~Bua, and R.~Mozzi.
\newblock Optical second-harmonic generation in crystals of organic dyes.
\newblock {\em App}, 15:393--396, 1969.

\bibitem{wayna00.01}
R.W. Waynant and M.N. Ediger.
\newblock {\em Electro-optics handbook}.
\newblock McGraw-Hill, 2000.

\bibitem{tutt93.01}
L.W. Tutt and T.F. Boggess.
\newblock A review of optical limiting mechanisms and devices using organics,
  fullerenes, semiconductors and other materials.
\newblock {\em Progress in quantum electronics}, 17(4):299--338, 1993.

\bibitem{yariv77.01}
A.~Yariv and D.M. Pepper.
\newblock Amplified reflection, phase conjugation, and oscillation in
  degenerate four-wave mixing.
\newblock {\em Optics letters}, 1(1):16--18, 1977.

\bibitem{yariv78.01}
A.~Yariv.
\newblock Phase conjugate optics and real-time holography.
\newblock {\em IEEE J. Quant. Elec.}, 14(9):650--660, 1978.

\bibitem{boyd92.01}
Robert~W. Boyd.
\newblock {\em Nonlinear Optics}.
\newblock Academic Press, 1992.

\bibitem{lytel86.02}
R.~Lytel.
\newblock Pump-depletion effects in noncollinear degenerate four-wave mixing in
  kerr media.
\newblock {\em J. Opt. Soc. Am. B}, 3:1580--1584, 1986.

\bibitem{winfu80.01}
H.G. Winful and J.H. Marburger.
\newblock Hysteresis and optical bistability in degenerate four-wave mixing.
\newblock {\em Applied Physics Letters}, 36:613, 1980.

\bibitem{gibbs84.01}
H.M. Gibbs.
\newblock Physics of optical bistability (a).
\newblock {\em J. Opt. Soc. Am. A, vol. 1, page 1281}, 1:1281, 1984.

\bibitem{weine11.01}
A.~Weiner.
\newblock {\em Ultrafast optics}, volume~72.
\newblock Wiley, 2011.

\bibitem{lytel84.01}
R.~Lytel.
\newblock Optical multistability in collinear degenerate four-wave mixing.
\newblock {\em J. Opt. Soc. Am. B}, 1:91--94, 1984.

\bibitem{vanec91.01}
TE~Van~Eck, AJ~Ticknor, RS~Lytel, and GF~Lipscomb.
\newblock Complementary optical tap fabricated in an electro-optic polymer
  waveguide.
\newblock {\em Appl Phys Lett}, 58(15):1588--1590, 1991.

\bibitem{zyss85.01}
J.~Zyss.
\newblock {Nonlinear Organic Materials for Intergrated Optics: A Review}.
\newblock {\em J. Mol. Electron.}, 1:25--45, 1985.

\bibitem{boyd09.01}
R.~W. Boyd.
\newblock {\em Nonlinear Optics}.
\newblock Academic Press, 3rd edition, 2009.

\bibitem{kuzyk10.01}
M.~G. Kuzyk.
\newblock A bird's-eye view of nonlinear-optical processes: Unification through
  scale invariance.
\newblock {\em Nonl. Opt. Quant. Opt.}, 40:1--13, 2010.

\bibitem{horna92.01}
L.~Hornak.
\newblock Polymers for lightwave and integrated optics: technology and
  applications.
\newblock {\em Marcel Dekker, Inc, 270 Madison Ave, New York, New York 10016,
  USA, 1992. 768}, 1992.

\bibitem{kuzyk06.06}
M.~G. Kuzyk.
\newblock {\em Polymer Fiber Optics: materials, physics, and applications},
  volume 117 of {\em Optical science and engineering}.
\newblock CRC Press, Boca Raton, 2006.

\bibitem{kuzyk00.01}
M.~G. Kuzyk.
\newblock {Physical Limits on Electronic Nonlinear Molecular Susceptibilities}.
\newblock {\em Phys. Rev. Lett.}, 85:1218, 2000.

\bibitem{kuzyk09.01}
M.~G. Kuzyk.
\newblock Using fundamental principles to understand and optimize
  nonlinear-optical materials.
\newblock {\em J. Mat. Chem.}, 19:7444--7465, 2009.

\bibitem{kuzyk06.03}
M.~G. Kuzyk.
\newblock {Fundamental limits of all nonlinear-optical phenomena that are
  representable by a second-order susceptibility}.
\newblock {\em J. Chem Phys.}, 125:154108, 2006.

\bibitem{kuzyk03.01}
M.~G. Kuzyk.
\newblock {Fundamental limits on third-order molecular susceptibilities:
  erratum}.
\newblock {\em Opt. Lett.}, 28:135, 2003.

\bibitem{zhou06.01}
J.~Zhou, M.~G. Kuzyk, and D.~S. Watkins.
\newblock {Pushing the hyperpolarizability to the limit}.
\newblock {\em Opt. Lett.}, 31:2891, 2006.

\bibitem{zhou07.02}
J.~Zhou, U.~B. Szafruga, D.~S. Watkins, and M.~G. Kuzyk.
\newblock {Optimizing potential energy functions for maximal intrinsic
  hyperpolarizability}.
\newblock {\em Phys. Rev. A}, 76:053831, 2007.

\bibitem{perez07.01}
J.~P\'{e}rez-Moreno, Y.~Zhao, K.~Clays, and M.~G. Kuzyk.
\newblock {Modulated conjugation as a means for attaining a record high
  intrinsic hyperpolarizability}.
\newblock {\em Opt. Lett.}, 32(1):59--61, 2007.

\bibitem{perez09.01}
J.~P\'{e}rez-Moreno, Y.~Zhao, K.~Clays, M.~G. Kuzyk, Y.~Shen, L.. Qiu, J.~Hao,
  and K.~Guo.
\newblock Modulated conjugation as a means of improving the intrinsic
  hyperpolarizability.
\newblock {\em J. Am. Chem. Soc.}, 131:5084--5093, 2009.

\bibitem{ather12.01}
TJ~Atherton, J.~Lesnefsky, GA~Wiggers, and RG~Petschek.
\newblock Maximizing the hyperpolarizability poorly determines the potential.
\newblock {\em J. Opt. Soc. Am. B}, 29:513--520, 2012.

\bibitem{burke12.01}
C.J. Burke, J.~Lesnefsky, R.G. Petschek, and T.J. Atherton.
\newblock Optimizing the second hyperpolarizability with minimally-parametrized
  potentials.
\newblock {\em arXiv preprint arXiv:1212.4069}, 2012.

\bibitem{bello08.01}
M.~Belloni and R.~W. Robinett.
\newblock Quantum mechanical sum rules for two model systems.
\newblock {\em Am. J. Phys.}, 76(9):798--806, 2008.

\bibitem{wang99.01}
S.~Wang.
\newblock Generalization of the thomas-reiche-kuhn and the bethe sum rules.
\newblock {\em Phys. Rev. A}, 60(1):262--266, 1999.

\bibitem{kuzyk06.01}
M.~G. Kuzyk.
\newblock {Truncated Sum Rules and their use in Calculating Fundamental Limits
  of Nonlinear Susceptibilities}.
\newblock {\em J. Nonl. Opt. Phys. \& Mat.}, 15(1):77--87, 2006.

\bibitem{kuzyk08.01}
M.~C. Kuzyk and M.~G. Kuzyk.
\newblock {Monte Carlo Studies of the Fundamental Limits of the Intrinsic
  Hyperpolarizability}.
\newblock {\em J. Opt. Soc. Am. B.}, 25(1):103--110, 2008.

\bibitem{shafe10.01}
S.~Shafei, M.~C. Kuzyk, and M.~G. Kuzyk.
\newblock Monte carlo studies of the intrinsic second hyperpolarizability.
\newblock {\em J. Opt. Soc Am. B}, 27:1849--1856, 2010.

\bibitem{watkins09.01}
D.~S. Watkins and M.~G. Kuzyk.
\newblock Optimizing the hyperpolarizability tensor using external
  electromagnetic fields and nuclear placement.
\newblock {\em J. Chem. Phys.}, 131:064110, 2009.

\bibitem{kotto97.01}
Tsampikos Kottos and Uzy Smilansky.
\newblock Quantum chaos on graphs.
\newblock {\em Phys. Rev. Lett.}, 79:4794--4797, Dec 1997.

\bibitem{kotto99.02}
T.~Kottos and U.~Smilansky.
\newblock Periodic orbit theory and spectral statistics for quantum graphs.
\newblock {\em Ann. Phys.}, 274(1):76--124, 1999.

\bibitem{blume02.01}
R.~Bl\"umel, Yu. Dabaghian, and R.~V. Jensen.
\newblock Explicitly solvable cases of one-dimensional quantum chaos.
\newblock {\em Phys. Rev. Lett.}, 88:044101, Jan 2002.

\bibitem{blume02.02}
R.~Bl\"umel, Y.~Dabaghian, and R.~V. Jensen.
\newblock Exact, convergent periodic-orbit expansions of individual energy
  eigenvalues of regular quantum graphs.
\newblock {\em Phys. Rev. E}, 65:046222, Apr 2002.

\bibitem{dabag04.01}
Yu. Dabaghian and R.~Bl\"umel.
\newblock Explicit spectral formulas for scaling quantum graphs.
\newblock {\em Phys. Rev. E}, 70:046206, Oct 2004.

\bibitem{dabag02.01}
Yu. Dabaghian, R.~Jensen, and R.~Bl{\"u}mel.
\newblock Spectra of regular quantum graphs.
\newblock {\em J. Exp. Theor. Phys.}, 94:1201--1215, 2002.
\newblock 10.1134/1.1493174.

\bibitem{dabag03.01}
Yu. Dabaghian and R.~Bl\"umel.
\newblock Explicit analytical solution for scaling quantum graphs.
\newblock {\em Phys. Rev. E}, 68:055201, Nov 2003.

\bibitem{dabag07.01}
Y.~Dabaghian.
\newblock Periodic orbit theory and the statistical analysis of scaling quantum
  graph spectra.
\newblock {\em Physical Review E}, 75(5):056214, 2007.

\bibitem{gnutz10.01}
S.~Gnutzmann, JP~Keating, and F.~Piotet.
\newblock Eigenfunction statistics on quantum graphs.
\newblock {\em Ann. Phys.}, 325(12):2595--2640, 2010.

\bibitem{lytel12.01}
Rick Lytel, Shoresh Shafei, and Mark~G Kuzyk.
\newblock Nonlinear optics of quantum graphs.
\newblock In {\em Proc. SPIE 8474}, pages 84740O--84740O--9, 2012.

\bibitem{shafe12.01}
S.~Shafei, R.~Lytel, and M.~G. Kuzyk.
\newblock Geometry-controlled nonlinear optical response of quantum graphs.
\newblock {\em J. Opt. Soc. Am. B}, 29:3419--3428, 2012.

\bibitem{lytel13.01}
Rick Lytel, Shoresh Shafei, Julian~H. Smith, and Mark~G. Kuzyk.
\newblock Influence of geometry and topology of quantum graphs on their
  nonlinear optical properties.
\newblock {\em Phys. Rev. A}, 87:043824, Apr 2013.

\bibitem{kuzyk00.02}
M.~G. Kuzyk.
\newblock {Fundamental limits on third-order molecular susceptibilities}.
\newblock {\em Opt. Lett.}, 25:1183, 2000.

\bibitem{jerph78.01}
J.~Jerphagnon, D.~Chemla, and R.~Bonneville.
\newblock The description of the physical properties of condensed matter using
  irreducible tensors.
\newblock {\em Adv. Phys.}, 27(4):609--650, 1978.

\bibitem{bance10.01}
T.~Bancewicz and Z.~O{\.z}go.
\newblock Irreducible spherical representation of some fourth-rank tensors.
\newblock {\em J. Comput. Meth. Sci. Eng.}, 10(3):129--138, 2010.

\bibitem{joffr92.01}
M.~Joffre, D.~Yaron, J.~Silbey, and J.~Zyss.
\newblock {Second Order Optical Nonlinearity in Octupolar Aromatic Systems}.
\newblock {\em J . Chem. Phys.}, 97(8):5607--5615, 1992.

\bibitem{pasto09.01}
ZS~Pastore and R.~Bl{\"u}mel.
\newblock An exact periodic-orbit formula for the energy levels of the
  three-pronged star graph.
\newblock {\em J. Phys. A}, 42:135102, 2009.

\bibitem{szafr10.01}
U.~B. Szafruga, M.~G. Kuzyk, and D.~S. Watkins.
\newblock Maximizing the hyperpolarizability of one-dimensional systems.
\newblock {\em J. Nonl. Opt. Phys. \& Mat.}, 19:379, 2010.

\end{thebibliography}

\end{document}